\documentclass[twocolumn,english,superscriptaddress,prx,amsmath,amssymb,floatfix,longbibliography,nofootinbib]{revtex4-2}
\usepackage[T1]{fontenc}
\usepackage[utf8]{inputenc}
\usepackage{color}
\usepackage{babel}
\usepackage{amsmath}
\usepackage{graphicx}
\usepackage{wasysym}
\usepackage{esint}
\PassOptionsToPackage{normalem}{ulem}
\usepackage{ulem}
\usepackage[unicode=true,pdfusetitle,
 bookmarks=true,bookmarksnumbered=false,bookmarksopen=false,
 breaklinks=false,pdfborder={0 0 0},pdfborderstyle={},backref=false,colorlinks=true]
 {hyperref}

\makeatletter

\usepackage{dcolumn}
\usepackage{bm}
\usepackage{color}
\usepackage{soul}
\usepackage{babel}
\usepackage{amsfonts}
\usepackage{slashed}
\usepackage{enumerate}
\usepackage{orcidlink}

\usepackage{babel}

\makeatother

\begin{document}
\global\long\def\sgn{\mathrm{sgn}}%
\global\long\def\ket#1{\left|#1\right\rangle }%
\global\long\def\bra#1{\left\langle #1\right|}%
\global\long\def\sp#1#2{\langle#1|#2\rangle}%
\global\long\def\abs#1{\left|#1\right|}%
\global\long\def\avg#1{\langle#1\rangle}%

\title{General theory of slow non-Hermitian evolution}
\author{Parveen Kumar\,\orcidlink{0000-0003-3132-203X}}
\affiliation{Department of Condensed Matter Physics, Weizmann Institute of Science,
Rehovot, 76100 Israel}
\affiliation{Department of Physics, Indian Institute of Technology Jammu, Jammu
181221, India}
\author{Yuval Gefen}
\affiliation{Department of Condensed Matter Physics, Weizmann Institute of Science,
Rehovot, 76100 Israel}
\author{Kyrylo Snizhko\,\orcidlink{0000-0002-7236-6779}}
\affiliation{Department of Condensed Matter Physics, Weizmann Institute of Science,
Rehovot, 76100 Israel}
\affiliation{Institute for Quantum Materials and Technologies, Karlsruhe Institute
of Technology, 76021 Karlsruhe, Germany}
\affiliation{Univ. Grenoble Alpes, CEA, Grenoble INP, IRIG, PHELIQS, 38000 Grenoble,
France}
\date{\today}
\begin{abstract}
Non-Hermitian systems are widespread in both classical and quantum
physics. The dynamics of such systems has recently become a focal
point of research, showcasing surprising behaviors that include apparent
violation of the adiabatic theorem and chiral topological conversion
related to encircling exceptional points (EPs). These have both fundamental
interest and potential practical applications. Yet the current literature
features a number of apparently irreconcilable results. Here we develop
a general theory for slow evolution of non-Hermitian systems and resolve
these contradictions. We prove an analog of the adiabatic theorem
for non-Hermitian systems and generalize it in the presence of\emph{
uncontrolled environmental fluctuations (noise}). The effect of noise
turns out to be crucial due to \emph{inherent exponential instabilities}
present in non-Hermitian systems. Disproving common wisdom, the end
state of the system is determined by the \emph{final} Hamiltonian
only, and is insensitive to other details of the evolution trajectory
in parameter space. Our quantitative theory, leading to transparent
physical intuition, is amenable to experimental tests. It provides
efficient tools to predict the outcome of the system's evolution,
avoiding the need to follow costly time-evolution simulations. Our
approach may be useful for designing devices based on non-Hermitian
physics and may stimulate analyses of classical and quantum non-Hermitian-Hamiltonian
dynamics, as well as that of quantum Lindbladian and hybrid-Liouvillian
systems.

\tableofcontents{}
\end{abstract}
\maketitle
~

\clearpage{}

\section{Background and the summary of results}

In recent years, non-Hermitian Hamiltonians have attracted significant
attention in the realms of both classical and quantum physics \citep{Bender2007,Bender2019a,Ashida2020,Bergholtz2021,Okuma2023,Banerjee2023}.
In classical physics, they naturally appear in the context of lossy
systems: friction in mechanical systems, resistive losses in electric
circuits, light absoprtion --- all these give rise to non-Hermitian
Hamiltonians underlying the system dynamics \citep{Bender2007,Bender2019a,Ashida2020}.
In quantum systems, they appear when particles or states have a finite
lifetime \citep{Levy1982,Naghiloo2019,Ashida2020,Michishita2020,Abbasi2020a,Chen2021,Chen2021a,Li2022a,Ohnmacht2024}
or if one considers measurements with postselected outcomes \citep{Ashida2020,Huang2019,Snizhko2020}.
Furthermore, quantum systems in contact with Markovian environment
can be described by noisy non-Hermitian Hamiltonians \citep{Chenu2017,Dupays2020,Dupays2021,Martinez-Azcona2023,Martinez-Azcona2024}
or by the Lindblad equation, whose Liouville superoperator can also
be written as a non-Hermitian matrix \citep{Avron2011a,Albert2016a,Fraas2017a,Minganti2019,Pick2019,Lieu2020,Kumar2021a,Kumar2021,Joye2022a,Bu2023,Zhou2023,Larson2023,Pavlov2024,Gao2025}.

Non-Hermitian systems give rise to new effects and behaviors that
do not appear in their Hermitian counterparts: e.g., non-Hermitian
skin effect \citep{Okuma2020,Kawabata2019,McDonald2020,Li2020f,Bergholtz2021,Okuma2023,Banerjee2023,Spring2023}
or topological phase transitions without gap closing \citep{Matsumoto2020}.
Capturing the phenomenology of non-Hermitian systems sometimes requires
adapting new unconventional methods, e.g., using a non-Bloch band
theory instead of the conventional --- and successful in Hermitian
systems --- Bloch theory \citep{Kawabata2020a}. The unusual properties
of non-Hermitian systems may assist in a number of applications such
as enhancing sensing sensitivity \citep{Wiersig2020b}, manipulating
quantum systems \citep{Li2022a,Teixeira2023}, creating polarization-locked
optical devices \citep{Li2022} and multimode switches \citep{Arkhipov2023,Arkhipov2025},
or improving performance of quantum algorithms \citep{Chen2022}.

In this context, one effect that has attracted much attention is non-Hermitian
state conversion. Under sufficiently slow non-Hermitian evolution,
any initial state of the system is converted to one specific eigenstate
of the Hamiltonian \citep{Uzdin2011,Berry2011,Gilary2013,Graefe2013,Milburn2015,Doppler2016,Hassan2017,Hassan2017b,Hassan2017c,Zhong2018,Zhang2018,Zhang2019d,Pick2019,Feilhauer2020,Li2020p,Choi2020,Ribeiro2021,Laha2021,Liu2020h,Sun2022,Chen2021a,Li2022,Nasari2022,Bu2023,Guria2023,Arkhipov2023}.
Particularly interesting is the case of closed trajectories, as one
can compare the effect of following the evolution in one or the opposite
direction. The behavior is then classified as either chiral (the coversion
result is different for the two directions) or non-chiral (the conversion
result is the same). The chirality has been associated with the mutual
location of the evolution trajectory in the parameter space and exceptional
points (EPs) \citep{Uzdin2011,Berry2011,Gilary2013,Milburn2015,Doppler2016,Hassan2017,Zhong2018,Pick2019,Choi2020,Ribeiro2021,Laha2021,Liu2020h,Chen2021a,Li2022,Guria2023,Bu2023,Fan2023a,Zhang2023b,Qi2024,Shan2024,Alizadeh2024,Zhang2024,Ho2024,Chavva2025}.
However, a number of numerical and experimental examples is known,
when this common intuition is challenged by numerical simulations
and experimental data \citep{Hassan2017b,Hassan2017c,Zhang2018,Zhang2019d,Sun2022,Nasari2022,Gao2025}.
This underlines the fact that analytical understanding of non-Hermitian
state conversion \citep{Berry2011,Uzdin2011,Gilary2013,Milburn2015,Hassan2017}
is limited and does not provide an explanation for the latter examples.

It is worthwhile mentioning that recent works \citep{Nye2023,Nye2024,Gao2025}
stressed the importance of non-adiabaticity effects for non-Hermitian
state conversion. References \citep{Nye2023,Nye2024} have shown that
the result of conversion crucially depends on the degree of adiabaticity
(evolution slowness) and that the conversion chirality does not qualitatively
depend on the proximity of an EP. In particular, notions of instantaneously
and trajectory-averaged dominant eigenstates have been introduced.
Notably, these works rely on exact solutions for specific evolution
examples --- rendering the generality of the conclusions unclear.

The present work tackles the problem of slow non-Hermitian evolution
from a general perspective. We develop a general theory of such dynamics
in the absence (“naïve theory”) and in the presence (“advanced theory”)
of uncontrolled fast perturbations (noise). We show that the inclusion
of \emph{noise is essential, and its presence drastically alters the
dynamics of non-Hermitian systems}. The crucial role played by noise
is a manifestation of \emph{inherent exponential instabilities} featured
by these systems.

While our theoretical analysis captures the general framework of slow
non-Hermitian evolution (both in the presence and absence of EPs),
we demonstrate its utility through the study of non-Hermitian state
conversion. We demonstrate that the chirality of the conversion indeed
has no relation to EPs. Instead, the system state is always converted
to the “end-point fastest growing” state (the instantaneously dominant
eigenstate at the end of the evolution trajectory). Furthermore, the
conversion chirality is fully determined by the behavior of the eigenvalues
of the Hamiltonian at the end points of the evolution trajectory.

Our results thus disprove the common wisdom, namely, that complete
analysis of spectral evolution throughout the trajecotry taken in
parameter space is needed in order to find the system's final state.
In fact, here we show that the slow evolution of such a system is
determined by its final point in parameter space, and is insensitive
to other characteristics of the trajectory.

Our quantitative theory leads to a transparent physical picture. Its
predictions are amenable to experimental tests. Our approach provides
efficient tools to theoretically study the slow evolution of the system,
avoiding the need to follow costly time-evolution simulations. Our
results can be straightforwardly generalized to quantum Lindbladian
or hybrid-Liouvillian dynamics.

The structure of the paper is as follows. In Sec.~\ref{subsec:Summary_Perfect-slow-evolution},
we present the general theory of noiseless slow non-Hermitian evolution
(``non-Hermitian adiabatic theorem''). We discuss how it predicts
the common-wisdom expectations for non-Hermitian state conversion.
We include noise into consideration in Sec.~\ref{subsec:Summary_Slow-evolution-with-fast-noise}
and demonstrate how it changes the predictions for non-Hermitian state
conversion. In Sec.~\ref{subsec:Summary_Implications}, we exemplify
how the predictions of the noiseless theory can fall wrong, while
the noise-aware theory always gives the correct prediction for numerical
simulations. We further demonstrate the theory's \emph{quantitative}
accuracy in Sec.~\ref{sec:quantitative_predictions_advanced_theory}.
In Sec.~\ref{sec:experimental_verification_proposals}, we discuss
how to verify our theory experimentally. We conclude with a brief
summary and outlook in Sec.~\ref{sec:Conclusions}.

We also provide several appendices. Appendix~\ref{app:Derivation_of_analytical_results}
describes the derivation of our analytical theory (both in the absence
and presence of noise). Appendix~\ref{app:Previous_explanations}
discusses the previous analytical explanations of unconventional conversion
behavior \citep{Hassan2017b,Hassan2017c,Zhang2018,Zhang2019d}: these
have never appealed to the presence uncontrolled perturbations, which
we show to be essential; we find most of these explanations erroneous,
while one \citep{Zhang2019d} is correct and consistent with our explanation,
yet limited in applicability. Appendix~\ref{app:quantitative_advanced_theory}
gives further numerical evidence for the quantitative accuracy of
our theory.

\section{Non-Hermitian adiabatic theorem (\textquotedblleft naïve theory\textquotedblright )}

\label{subsec:Summary_Perfect-slow-evolution}

We consider the problem of evolution under a time-dependent non-Hermitian
Hamiltonian $H(t)=H_{s}$ with $s=t/T\in[0,1]$, having in mind the
limit of slow evolution, $T\rightarrow\infty$. The key object of
interest is the evolution operator
\begin{equation}
\mathcal{U}(T,0)=\mathcal{T}\exp\left(-i\int_{0}^{T}dt\,H(t)\right),\label{eq:evolution_operator_definition}
\end{equation}
where $\mathcal{T}$ stands for time ordering. Of particular interest
is the effect of the evolution operator on the instantaneous eigenbasis
of $H(t)$. To this end, we define the instantaneous eigenbasis through
diagonalizing the Hamiltonian:
\begin{equation}
H(t)\equiv H_{s}=U(t)D(t)U^{-1}(t)\equiv U_{s}D_{s}U_{s}^{-1}.\label{eq:H_diagonalization}
\end{equation}
The matrix $D(t)$ is composed of the Hamiltonian eigenvalues $\lambda_{n}(t)\equiv\lambda_{n,s}$
standing on the diagonal. The matrix $U(t)\equiv U_{s}$ encodes the
instantaneous right eigenvectors of the Hamiltonian, $\ket{n(t)}=\ket{n_{s}}=U_{s}\ket n$,
where $\ket n$ is a column vector $\left(0,\ldots,0,1,0,\ldots0\right)^{\mathrm{T}}$
with $1$ on the $n$th postition. Note that we have implicitly assumed
that such diagonalization is possible; we will further assume that
it is unique. That is, we assume that there are no exceptional or
degeneracy points \emph{on the evolution trajectory} (these may be
present elsewhere in the space of control parameters).

We develop a systematic expansion in powers of $T^{-1}$ and show
that it is possible to represent the evolution operator as
\begin{equation}
\mathcal{U}(T,0)=U(T)\mathcal{E}(T,0)U^{-1}(0),\label{eq:summary_perfectly_slow_evolution_1}
\end{equation}
\begin{multline}
\mathcal{E}(T,0)=\left(\mathbb{I}+W(T)\right)\\
\times\exp\left(-i\int_{0}^{T}dt\,D(t)\right)\left(\mathbb{I}+\tilde{W}(0)\right).\label{eq:summary_perfectly_slow_evolution_2}
\end{multline}
Here $\mathcal{E}(T,0)$ encodes the evolution operator's action in
the instantaneous eigenbasis. The operators $W(t)$ and $\tilde{W}(t)$
are both $O(T^{-1})$ and are related to the eigenbasis evolution
operator, $U^{-1}(t)\partial_{t}U(t)\propto T^{-1}$. Finally, we
have neglected corrections $O(T^{-1})$ to the Hamiltonian eigenvalues,
$D(t)$, inside the exponential --- these are unimportant for our
purposes. The derivation, detailed expressions for $W(t)$, $\tilde{W}(t)$,
and the omitted corrections to $D(t)$, as well as a discussion of
a few subtleties, can be found in Appendix~\ref{subsec:Iterative-adiabatic-expansion}.

We call equations (\ref{eq:summary_perfectly_slow_evolution_1}--\ref{eq:summary_perfectly_slow_evolution_2})
the adiabatic theorem for non-Hermitian Hamiltonians. One sees that
it is consistent with the common wisdom about non-Hermitian state
conversion. Instantaneous eigenstates of the Hamiltonian grow/decay
under the Hamiltonian's action, as characterized by $\mathrm{Im}\,\int_{0}^{T}dt\,\lambda_{n}(t)$;
the state with the largest value of this integral dominates the evolution
by its end --- which is why essentially any initial state\footnote{Strictly speaking, if for some initial state, $\ket{\psi_{in}}$,
$\left(\mathbb{I}+\tilde{W}(0)\right)\ket{\psi_{in}}$ does not include
a certain instantaneous eigenstate $\ket{n(t=0)}$, the adiabatically-time-evolved
descendant of this eigenstate will not be generated and will not be
amplified. In particular, if the most growing state is absent, the
conversion to it will not take place. This is, however, a highly fine-tuned
situation. We refer to any state except for such highly fine-tuned
ones as ``essentially any state''.} gets converted to this ``\emph{most growing}'' state.

To our knowledge, our result is the first proof of this conversion
property of slow non-Hermitian evolution in complete generality, beyond
previous proofs for two-level systems \citep{Uzdin2011,Gilary2013,Milburn2015,Nye2024}.
Note also that the conversion is only accurate up to $O(T^{-1})$,
as implied by $W(T)$, and not exponentially accurate in $T$. This
has been previously pointed out in Ref.~\citep{Ribeiro2021}.

We note in passing the substantial mathematical literature on adiabatic
evolution with non-Hermitian Hamiltonians \citep{Kvitsinsky1991a,Nenciu1992a,Avron2012a},
Hermitian Hamiltonians in complex time \citep{Hwang1977a,Joye1991b,Joye1991c,Joye1997a,Joye2007a},
or Lindbladians \citep{Avron2011a,Fraas2017a,Joye2022a}. These works
focused mostly on the dynamics within the subspace of the states with
the largest $\mathrm{Im}\,\lambda_{n}(t)$ \citep{Nenciu1992a,Joye2007a,Avron2012a},
as well as on Landau-Zener transition probabilities \citep{Hwang1977a,Kvitsinsky1991a,Joye1991b,Joye1991c,Joye1997a,Avron2011a,Fraas2017a,Joye2022a}.
This is distinct from our focus here, namely, conversion between eigenstates
with different $\mathrm{Im}\,\lambda_{n}(t)$, for which Eqs.~(\ref{eq:summary_perfectly_slow_evolution_1}--\ref{eq:summary_perfectly_slow_evolution_2})
provide an explicit answer.

\section{Slow non-Hermitian evolution in the presence of noise (\textquotedblleft advanced
theory\textquotedblright )}

\label{subsec:Summary_Slow-evolution-with-fast-noise}

The consideration above assumed perfectly-controlled slow evolution.
However, one cannot expect this in reality. In experiments, noise
and uncontrolled perturbations will lead to appearance of an additional
term in the Hamiltonian $H(t)\rightarrow\bar{H}(t)=H(t)+\varepsilon\delta H(t)$.
In numerical simulations of non-Hermitian dynamics, numerical errors
may and will occur. These can similarly be modelled by $\delta H(t)$.
While both are typically small, we are dealing with a system that
has exponential instabilities. Therefore, incorporating the analysis
of perturbations is essential.

Our goal is to calculate the evolution operator in the presence of
perturbations:
\begin{equation}
\mathcal{\bar{U}}(T,0)=\mathcal{T}\exp\left(-i\int_{0}^{T}dt\,\bar{H}(t)\right).
\end{equation}
If the perturbation $\delta H(t)$ is also slow, the analysis of Sec.~\ref{subsec:Summary_Perfect-slow-evolution}
can be repeated with the modified Hamiltonian, leading to a small
modification of the instantaneous eigenstates and eigenvalues, but
no qualitative change. Yet there is no reason to expect slowness from
uncontrolled perturbations.

For fast perturbations, we expand $\mathcal{\bar{U}}(T,0)$ in perturbation
series:
\begin{multline}
\mathcal{\bar{U}}(T,0)=\mathcal{U}(T,0)\\
-i\varepsilon\int_{0}^{T}dt_{1}\mathcal{U}(T,t_{1})\delta H(t_{1})\mathcal{U}(t_{1},0)\\
+O\left(\varepsilon^{2}\right),\label{eq:summary_evolution_operator_deltaH_perturbation_theory_first_order}
\end{multline}
where
\begin{equation}
\mathcal{U}(t_{2},t_{1})=\mathcal{T}\exp\left(-i\int_{t_{1}}^{t_{2}}dt\,H(t)\right)
\end{equation}
 represents unperturbed (noiseless) slow evolution and can be expressed
in the form (\ref{eq:summary_perfectly_slow_evolution_1}--\ref{eq:summary_perfectly_slow_evolution_2})
with appropriately replaced time limits.

Note that $\mathcal{U}(T,0)=\mathcal{U}(T,t_{1})\mathcal{U}(t_{1},0)$
for any $t_{1}$, so that the role of $\delta H(t_{1})$ is to disrupt
the noiseless adiabatic evolution at all possible intermediate times.
Focus on $\delta H(t_{1})\mathcal{U}(t_{1},0)$. By the time $t_{1}$,
essentially any initial state becomes aligned with one specific eigenstate
$\ket{n(t_{1})}$ due to the unperturbed evolution $\mathcal{U}(t_{1},0)$.
The perturbation at time $t_{1}$ generates amplitude of order $\varepsilon$
in other (generically --- all other) instantaneous eigenstates of
$H(t_{1})$. This effectively restarts the non-Hermitian state conversion
process --- with a new initial state and a different evolution trajectory:
$H(t\in[t_{1},T])$ instead of $H(t\in[0,T])$. If the preferred state
of $\mathcal{U}(T,t_{1})$ for some $t_{1}$ is different from that
of $\mathcal{U}(T,0)$, the evolution outcome will change.

Namely, all states will be converted not to the most growing state
(with the largest $\mathrm{Im}\int_{0}^{T}dt\,\lambda_{n}(t)$), but
to the ``\emph{end-point fastest growing}'' one --- the one that
has the largest $\mathrm{Im}\,\lambda_{n}(t)$ for the times near
the end of the evolution (that is, for $t\in[T(1-y),T]$ for some
interval length $yT$). One can understand this from the following
consideration: all the ``restarts'' of conversion at $t_{1}\in[T(1-y),T]$
lead to the same the evolution outcome --- conversion to the end-point
fastest growing state. We illustrate this point with several numerical
examples in Sec.~\ref{subsec:Summary_Implications}.

\section{Implications for non-Hermitian state conversion}

\label{subsec:Summary_Implications}

Here we illustrate the predictions of the general theory above with
a few examples using a specific $2\times2$ non-Hermitian Hamiltonian.
We discuss example trajectories that highlight the key effects predicted
by our theory.

We use the Hamiltonian
\begin{equation}
H=\begin{pmatrix}\delta(t)+ig(t) & -1\\
-1 & -\delta(t)-ig(t)
\end{pmatrix}.\label{eq:nHH_examples}
\end{equation}
For each value of real parameters $\delta$ and $g$, it has two eigenvalues,
$\lambda_{\pm}=\pm\sqrt{1+\left(\delta+ig\right)^{2}}$, and the corresponding
right eigenstates $\ket{\psi_{\pm}}$. The Hamiltonian possesses two
EPs: $\left(\delta=0,g=\pm1\right)$. We will consider the system
dynamics as the Hamiltonian parameters change according to
\begin{align}
\delta(t)= & \delta_{0}-R\sin\left(\omega t+\phi\right),\nonumber \\
g(t)= & g_{0}-R\cos\left(\omega t+\phi\right),\label{eq:trajectory_general}
\end{align}
where parameters $\delta_{0}$, $g_{0}$, $R$, $\phi$ determine
the specific trajectory as $t\in[0,T]$. The rate of evolution, $\omega$,
is always chosen such that $\abs{\omega}\ll\abs{\lambda_{+}(t)-\lambda_{-}(t)}$,
which guarantees the applicability of the above theory for \emph{slow}
evolution.

In the course of the evolution, the system state
\begin{equation}
\ket{\psi(t)}=\mathcal{N}(t)\left(c_{+}(t)\ket{\psi_{+}(t)}+c_{-}(t)\ket{\psi_{-}(t)}\right),
\end{equation}
where $\mathcal{N}(t)$ is the normalization constant chosen in such
a way that $\abs{c_{+}(t)}^{2}+\abs{c_{-}(t)}^{2}=1$. Throughout
the paper, we call $\abs{c_{\pm}(t)}^{2}$ the population of the respective
instantaneous eigenstate.

According the theory of Sec.~\ref{subsec:Summary_Perfect-slow-evolution}
and Eqs.~(\ref{eq:summary_perfectly_slow_evolution_1}--\ref{eq:summary_perfectly_slow_evolution_2}),
$c_{\pm}(t)\propto\exp\left(\mathrm{Im}\int_{0}^{t}dt\,\lambda_{\pm}(t)\right)$
(modulo the small corrections arising from $\tilde{W}(0)$ and $W(T)$).
This dictates non-Hermitian conversion into the most growing state:
the one with the biggest value of $\mathrm{Im}\int_{0}^{T}dt\,\lambda(t)$,
when integrated over the whole trajectory. We will refer to this prediction
as the naïve theory.

At the same time, Sec.~\ref{subsec:Summary_Slow-evolution-with-fast-noise}
and Eq.~(\ref{eq:summary_evolution_operator_deltaH_perturbation_theory_first_order})
predict a more involved dynamics, ultimately leading to the conversion
to the eigenstate that has the biggest $\mathrm{Im}\,\lambda(t)$
for the times towards the end of the trajectory. That is, to the end-point
fastest growing state. We will refer to this prediction as the advanced
theory.

Below we illustrate that the predictions of the advanced theory can
always be trusted, whereas the naïve theory can make errors in predicting
the result of conversion. We first illustrate this for open trajectories
in Sec.~\ref{subsec:summary_most_vs_last_growing}. Then we focus
on closed trajectories and illustrate how the chirality of non-Hermitian
state conversion becomes disentangled from encircling EPs, cf.~Secs.~\ref{subsec:non-chiral-with-encircling}
and \ref{subsec:chiral-without-encircling}. These examples illustrate
the \emph{qualitative} predictive power of the advanced theory. For
a demonstration that the advanced theory gives \emph{quantitatively}
accurate predictions, we refer the reader to Sec.~\ref{sec:quantitative_predictions_advanced_theory}.

\begin{figure}[t]
\begin{centering}
\includegraphics[width=1\columnwidth]{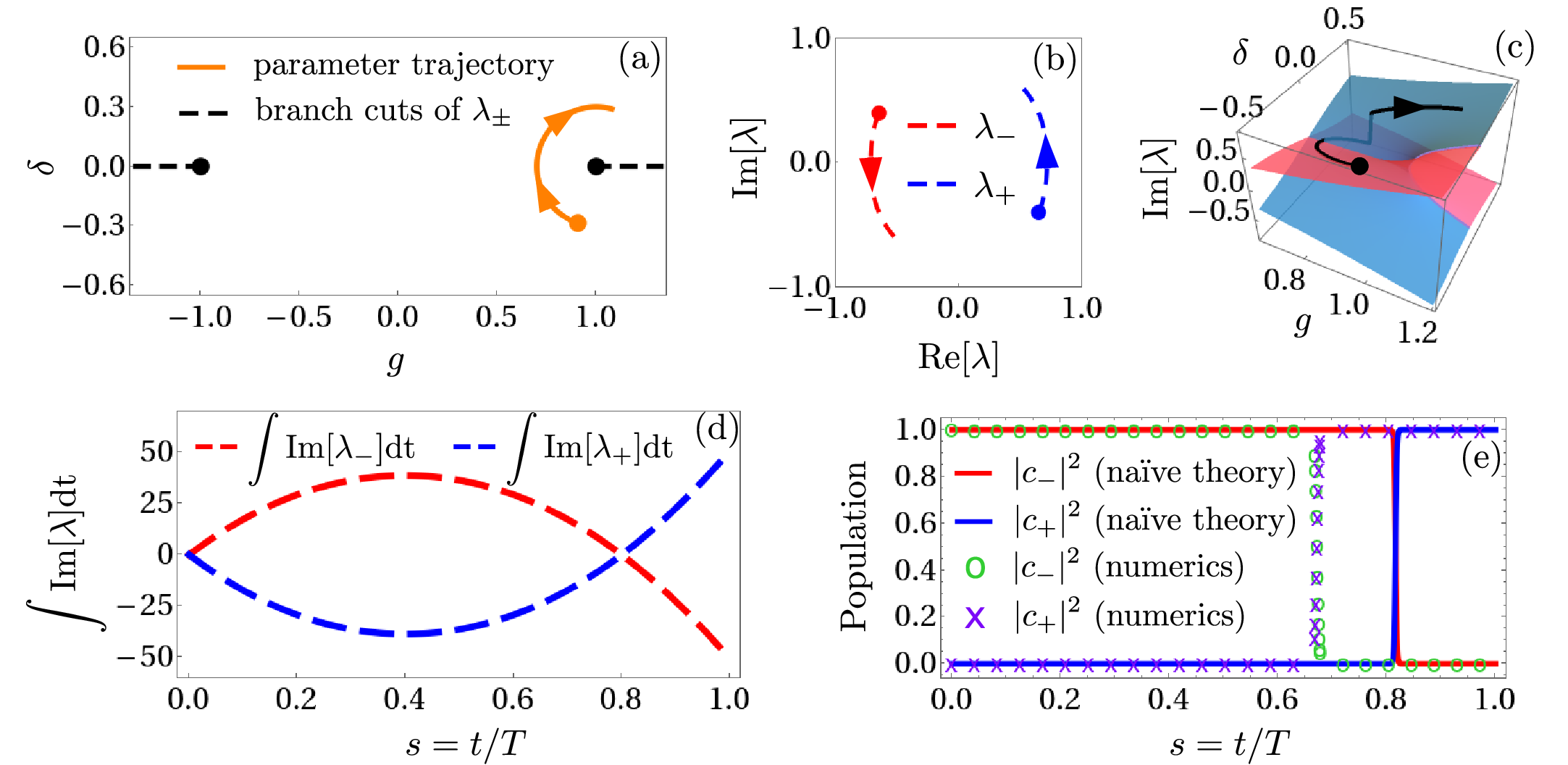}
\par\end{centering}
\caption{\textbf{Open-trajectory evolution under Hamiltonian (\ref{eq:nHH_examples}).
The most growing state wins, in qualitative agreement with the naïve
theory of Sec.~\ref{subsec:Summary_Perfect-slow-evolution}. This
also agrees with the advanced theory of Sec.~\ref{subsec:Summary_Slow-evolution-with-fast-noise},
as the most growing state is also the end-point fastest growing one.}
(a) The trajectory in the parameter plane (orange) where the trajectory
direction is shown with arrows and the orange dot depicts the starting
point. The trajectory corresponds to the following parameters in Eq.~(\ref{eq:trajectory_general}):
$\delta_{0}=0$, $g_{0}=1$, $R=0.3$, $T=500$, $\omega=-\pi/T$,
$\phi=0.4\pi$. The black dots correspond to the EPs of the Hamiltonian,
and the dashed lines correspond to the branch cuts of $\lambda_{\pm}$.
(b) The trajectory of the eigenvalues $\lambda_{\pm}$ with arrows
showing the trajectory direction. The eigevalues $\lambda_{\pm}$
at the start of the parameter trajectory are depicted as blue and
red dots. (c) Riemann surface plot of $\mathrm{Im}\,\lambda_{\pm}(t)$
(red surface for $\lambda_{-}$, blue surface for $\lambda_{+}$).
The state trajectory (black), with black dot depicting the starting
point and the arrow showing the trajectory direction, corresponds
to $\protect\abs{c_{+}(t)}^{2}\mathrm{Im}\,\lambda_{+}(t)+\protect\abs{c_{-}(t)}^{2}\mathrm{Im}\,\lambda_{-}(t)$.
(d) The integrals $\mathrm{Im}\int_{0}^{t}dt\,\lambda_{\pm}(t)$ that
govern the non-Hermitian state conversion under the naïve theory.
(e) Population dynamics: the prediction of the naïve theory vs the
numerical simulation. While qualitatively they agree --- the state
is ultimately converted to $\protect\ket{\psi_{+}}$ --- the location
of the population switch is not the same ( $t_{1}\approx0.8T$ vs
$t_{1}'\apprle0.7T$).}

\label{fig:open_trajectory_no_surprises}
\end{figure}

\begin{figure}[t]
\centering{}\includegraphics[width=1\columnwidth]{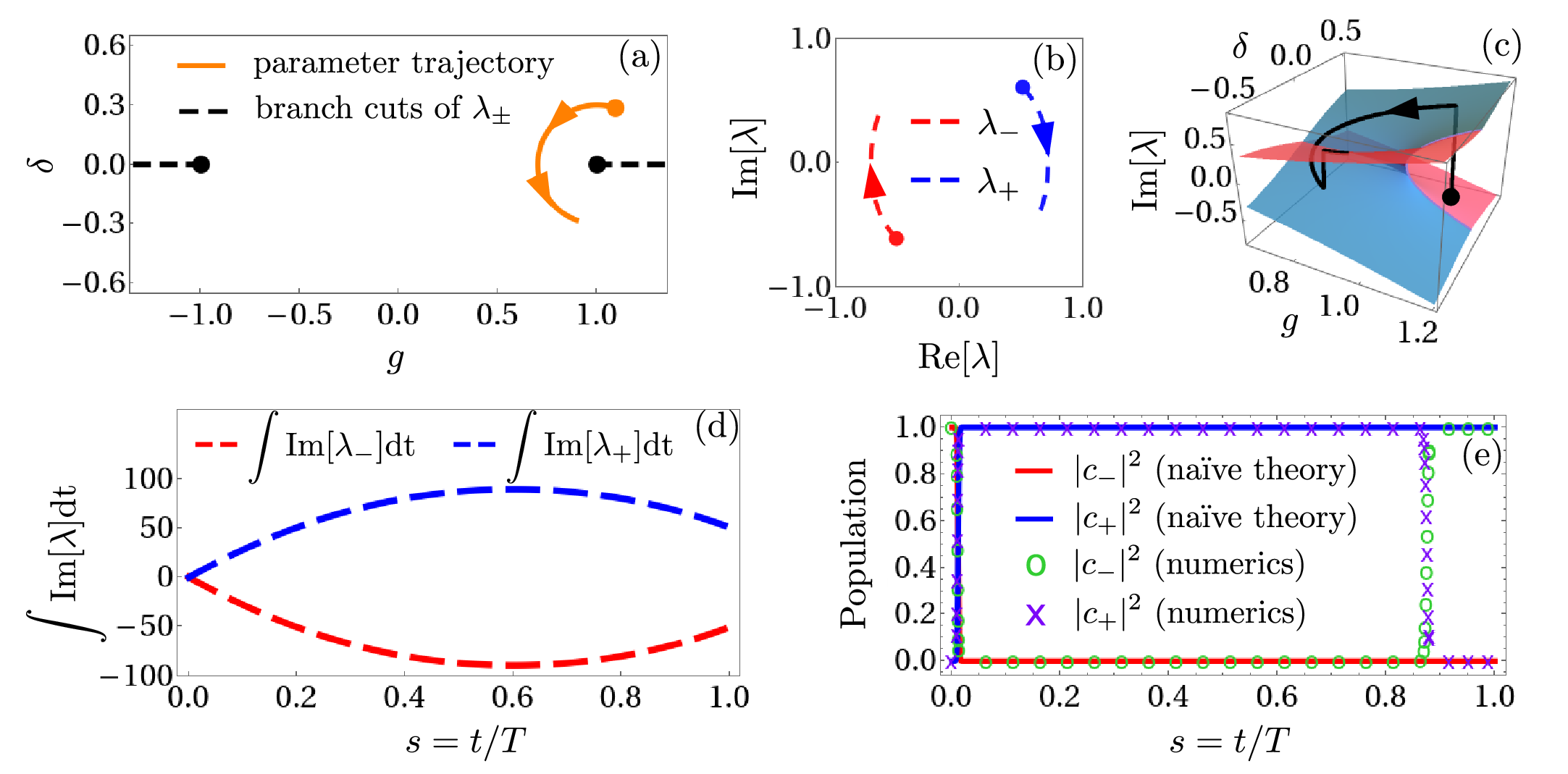}\caption{\label{fig:open_trajectory_qualitative_surprise}\textbf{ Open-trajectory
evolution under Hamiltonian (\ref{eq:nHH_examples}). The most growing
state loses, in disagreement with the naïve theory of Sec.~\ref{subsec:Summary_Perfect-slow-evolution}.
The end-point fastest growing state wins, in agreement with the advanced
theory of Sec.~\ref{subsec:Summary_Slow-evolution-with-fast-noise},
which accounts for the errors of numerical simulation (in this example)
and Hamiltonian control (in real experiments).} (a) The trajectory
in the parameter plane (orange) where the trajectory direction is
shown with arrows and the orange dot depict the starting point. The
trajectory corresponds to the following parameters in Eq.~(\ref{eq:trajectory_general}):
$\delta_{0}=0$, $g_{0}=1$, $R=0.3$, $T=500$, $\omega=\pi/T$,
$\phi=-0.6\pi$. The black dots correspond to the EPs of the Hamiltonian,
and the dashed lines correspond to the branch cut of $\lambda_{\pm}$.
(b) The trajectory of the eigenvalues $\lambda_{\pm}$ with arrows
showing the trajectory direction. The eigevalues $\lambda_{\pm}$
at the start of the parameter trajectory are depicted as blue and
red dots. (c) Riemann surface plot of $\mathrm{Im}\,\lambda_{\pm}(t)$
(red surface for $\lambda_{-}$, blue surface for $\lambda_{+}$).
The state trajectory (black) , with black dot depicting the starting
point and the arrow showing the trajectory direction, corresponds
to $\protect\abs{c_{+}(t)}^{2}\mathrm{Im}\,\lambda_{+}(t)+\protect\abs{c_{-}(t)}^{2}\mathrm{Im}\,\lambda_{-}(t)$.
(d) The integrals $\mathrm{Im}\int_{0}^{t}dt\,\lambda_{\pm}(t)$ that
govern the non-Hermitian state conversion under the naïve theory.
(e) Population dynamics: the prediction of the naïve theory vs the
numerical simulation. The naïve theory predicts conversion to the
most growing state $\protect\ket{\psi_{+}}$, whereas the numerical
simulation results in conversion to $\protect\ket{\psi_{-}}$. The
latter is in agreement with the advanced theory, as $\protect\ket{\psi_{-}}$
is the end-point fastest growing state.}
\end{figure}

\subsection{\label{subsec:summary_most_vs_last_growing}Which state wins: the
\uline{most growing} vs the \uline{end-point fastest growing}}

We now discuss examples of how the naïve theory fails in quantitative
(Fig.~\ref{fig:open_trajectory_no_surprises}) and, most importantly,
qualitative (Fig.~\ref{fig:open_trajectory_qualitative_surprise})
predictions for slow non-Hermitian evolution. We explain how the advanced
theory gives the correct predictions.

Consider an open trajectory in the parameter space, cf.~Fig.~\ref{fig:open_trajectory_no_surprises}(a).
Figure~\ref{fig:open_trajectory_no_surprises}(b) shows the evolution
of real and imaginary parts of $\lambda_{\pm}$ along the trajectory.
As explained in Secs.~\ref{subsec:Summary_Perfect-slow-evolution},
\ref{subsec:Summary_Slow-evolution-with-fast-noise}, the imaginary
part $\mathrm{Im}\,\lambda_{\pm}$ is of key importance. Fig.~\ref{fig:open_trajectory_no_surprises}(c)
shows the state dynamics superimposed with the Riemann surface of
$\lambda_{\pm}$. Namely, the black line shows $\abs{c_{+}(t)}^{2}\mathrm{Im}\,\lambda_{+}(t)+\abs{c_{-}(t)}^{2}\mathrm{Im}\,\lambda_{-}(t)$.
One can see the typical feature of non-Hermitian state converison:
starting in $\ket{\psi_{-}}$ with $\mathrm{Im}\,\lambda_{-}>0$,
the system state adiabatically follows $\ket{\psi_{-}}$ until it
suddenly switches to $\ket{\psi_{+}}$. The switch happens not immediately
at the point at which $\mathrm{Im}\,\lambda_{-}=\mathrm{Im}\,\lambda_{+}=0$
(after which $\mathrm{Im}\,\lambda_{-}<0$ and $\mathrm{Im}\,\lambda_{+}>0$),
but sometime later.

This can be qualitatively explained with the naïve theory of Sec.~\ref{subsec:Summary_Perfect-slow-evolution},
Eqs.~(\ref{eq:summary_perfectly_slow_evolution_1}--\ref{eq:summary_perfectly_slow_evolution_2}).
Indeed, Fig.~\ref{fig:open_trajectory_no_surprises}(d) shows the
evolution of $\int_{0}^{t}dt\,\mathrm{Im}\,\lambda_{\pm}(t)$. The
integrals change their trends (growing/decreasing) at $t_{0}\approx0.4T$,
where $\mathrm{Im}\,\lambda_{-}=\mathrm{Im}\,\lambda_{+}=0$, however
the amplification factors $\exp\left(\mathrm{Im}\int_{0}^{t}dt\,\lambda_{\pm}(t)\right)$
for the two states will only become equal at $t_{1}\approx0.8T$,
and soon after that, the population switch to $\ket{\psi_{+}}$ will
happen.

The naïve theory, however, fails to explain the dynamics quantitatively,
cf.~Fig.~\ref{fig:open_trajectory_no_surprises}(e). The numerical
simulation shows that the population switch between $\ket{\psi_{-}}$
and $\ket{\psi_{+}}$ takes place at $t_{1}'\apprle0.7T$, i.e., significantly
earlier than $\int_{0}^{t}dt\,\mathrm{Im}\,\lambda_{-}(t)=\int_{0}^{t}dt\,\mathrm{Im}\,\lambda_{+}(t)$.
This can be explained by the advanced theory of Sec.~\ref{subsec:Summary_Perfect-slow-evolution}:
The numerical simulation involes Trotterization of the evolution,
which means that there is inherent error introduced every numerical
time step $dt$. Interpreting this error as $\delta H(t)$ in Eq.~(\ref{eq:summary_evolution_operator_deltaH_perturbation_theory_first_order}),
one can say that the error creates a small but finite population in
$\ket{\psi_{+}}$. Similarly to the naïve theory, the population of
$\ket{\psi_{+}}$ gets amplified once $t>t_{0}$. However, unlike
in the naïve theory, the newly-created population of $\ket{\psi_{+}}$
does not need to overcome the suppression factor $\exp\left(\mathrm{Im}\int_{0}^{t_{0}}dt\,\left[\lambda_{+}(t)-\lambda_{-}(t)\right]\right)$
accumulated previously. This leads to the populations of $\ket{\psi_{+}}$
and $\ket{\psi_{-}}$ becoming equal at $t_{1}'<t_{1}$.

A much more striking distinction between the predictions of the naïve
and the advanced theories appears when one follows the same parameter
trajectory in the reverse direction, cf.~Fig.~\ref{fig:open_trajectory_qualitative_surprise}.
Here the final state happens to be $\ket{\psi_{-}}$, and not $\ket{\psi_{+}}$,
in contradiction with the naïve theory, cf.~Fig.~\ref{fig:open_trajectory_qualitative_surprise}(e).
This is, however, in agreement with the advanced theory, which predicts
conversion to the end-point fastest growing state, which is $\ket{\psi_{-}}$:
for $t_{0}\gtrsim0.6T$, $\mathrm{Im}\,\lambda_{-}>0>\mathrm{Im}\,\lambda_{+}$,
cf.~Figs.~\ref{fig:open_trajectory_qualitative_surprise}(b, d).
The mechanism is, again, the same: due to numerical errors, at every
time step, a small fraction of population of $\ket{\psi_{+}}$ is
converted to $\ket{\psi_{-}}$, whose population grows once $t_{0}\gtrsim0.6T$;
unlike in the naïve theory, the amplification factor does not need
to overcome the huge suppression accumulated previously, $\exp\left(\mathrm{Im}\int_{0}^{t_{0}}dt\,\left[\lambda_{-}(t)-\lambda_{+}(t)\right]\right)$.

We strongly emphasize the following. One might think that numerical
errors are an artifact of the simulation. That they can be eliminated
with better numerical schemes and do not reflect experimental reality.
In fact, quite the opposite stance should be taken. In a system with
exponential instabilities, errors (however small) can lead to a qualitative
change of behavior. Therefore, a better numerical scheme that reduces
the uncontrolled errors, will not eliminate the reported effects as
long as the evolution is sufficiently slow. That is, as long as the
instabilities have enough time to grow and manifest. Similarly, any
experimental implementation will necessarily have noise in the control
parameters (even if tiny). Therefore, the behavior illustrated above
--- non-Hermitian evolution brings the system to the end-point fastest
growing state, and not to the most growing one --- is not an artifact,
but is to be generically expected in the limit of slow evolution ($T\rightarrow\infty$).

\subsection{\label{subsec:non-chiral-with-encircling}Non-chiral state conversion
upon encircling the exceptional point}

\begin{figure*}[p]
\begin{centering}
\includegraphics[width=1\textwidth]{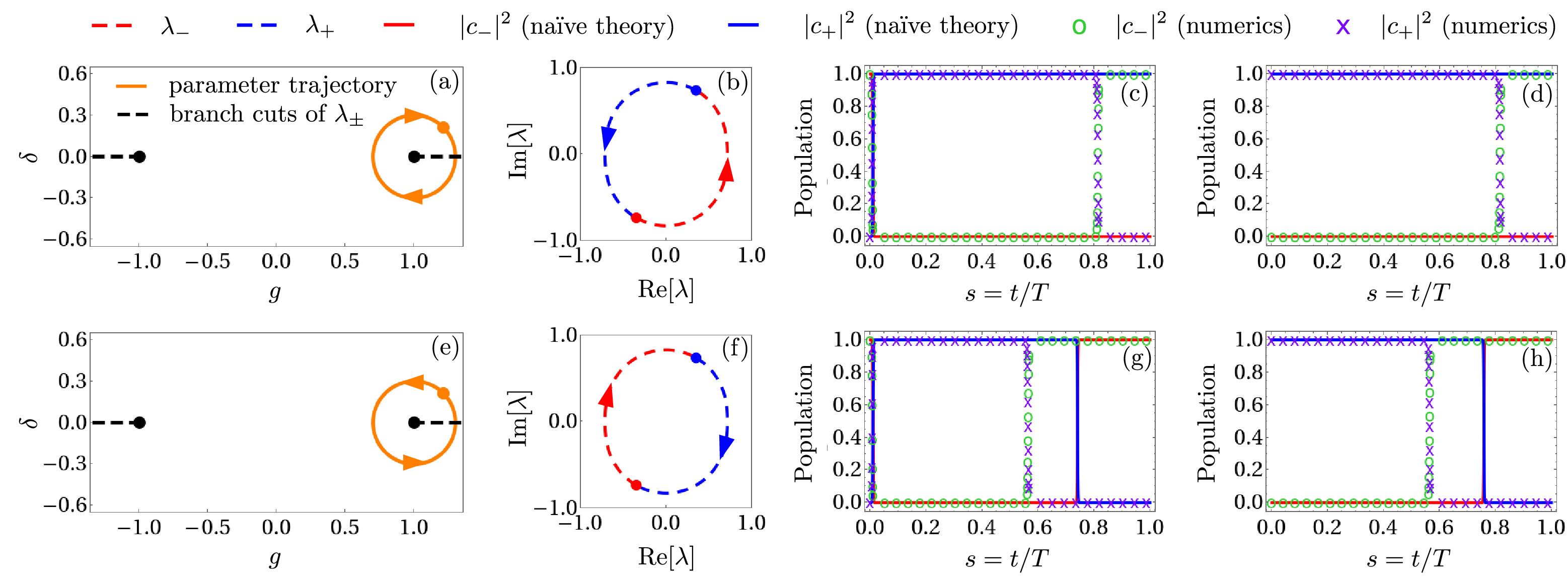}
\par\end{centering}
\caption{\textbf{Non-chiral state conversion when encircling an exceptional
point. The numerical result is in clear discrepancy with the predictions
of the naïve theory. The advanced theory explains the numerical result,
as $\protect\ket{\psi_{-}}$ is the end-point fastest growing state
for both encircling directions.} (a) The trajectory for clockwise
encircling in the parameter plane (orange) corresponds to the following
parameters in Eq.~(\ref{eq:trajectory_general}): $\delta_{0}=0$,
$g_{0}=1$, $R=0.3$, $T=500$, $\omega=-2\pi/T$, $\phi=-3\pi/4$.
The trajectory direction is shown with arrows and the orange dot depict
the starting point. The black dots correspond to the EPs of the Hamiltonian,
and the dashed lines correspond to the branch cuts of $\lambda_{\pm}$.
(b) The trajectory of the eigenvalues $\lambda_{\pm}$ corresponding
to the clockwise trajectory. The eigevalues $\lambda_{\pm}$ at the
start of the parameter trajectory are depicted as blue and red dots,
and the arrows show the trajectory direction. (c) The population dynamics
for clockwise encircling when the system is initialized in $\protect\ket{\psi_{-}}$.
(d) The population dynamics for clockwise encircling when the system
is initialized in $\protect\ket{\psi_{+}}$. (e) The trajectory for
counterclockwise encircling in the parameter plane (orange) corresponds
to the following parameters in Eq.~(\ref{eq:trajectory_general}):
$\delta_{0}=0$, $g_{0}=1$, $R=0.3$, $T=500$, $\omega=2\pi/T$,
$\varphi=-3\pi/4$. (f,g,h) Same as (b,c,d), but for the counterclockwise
trajectory.}

\label{fig:non-chiral-with-encircling}
\end{figure*}
\begin{figure*}[p]
\begin{centering}
\includegraphics[width=1\textwidth]{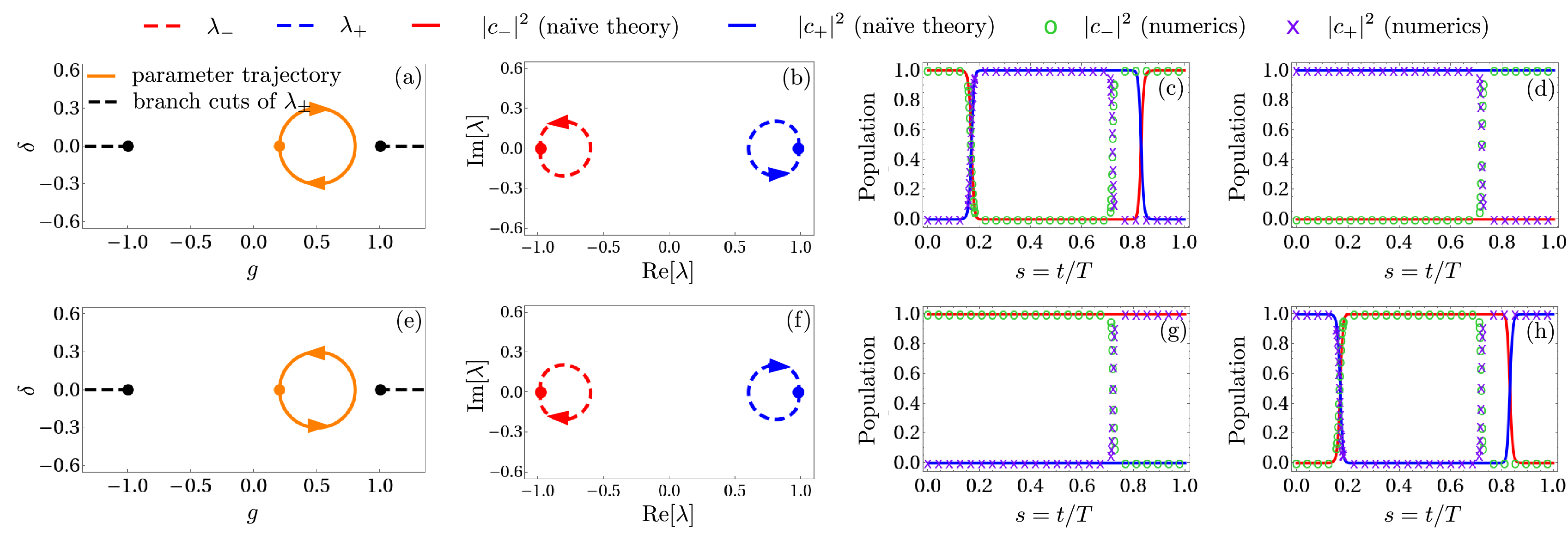}
\par\end{centering}
\caption{\textbf{Chiral state conversion without encircling an exceptional
point. The naïve theory predicts no conversion whatsoever, while the
numerics predicts conversion to $\protect\ket{\psi_{-}}$ ($\protect\ket{\psi_{+}}$)
for clockwise (counterclockwise) trajectory. This is explained by
the advanced theory as the respective states are the end-point fastest
growing in the respective cases.} (a) The clockwise trajectory (orange)
corresponds to the following parameters in Eq.~(\ref{eq:trajectory_general}):
$\delta_{0}=0$, $g_{0}=0.5$, $R=0.3$, $T=500$, $\omega=-2\pi/T$,
$\phi=0$. The trajectory direction is shown with arrows and the orange
dot depict the starting point. The black dots correspond to the EPs
of the Hamiltonian, and the dashed lines correspond to the branch
cuts of $\lambda_{\pm}$. (b) The trajectory of the eigenvalues $\lambda_{\pm}$
corresponding to the clockwise trajectory. The eigevalues $\lambda_{\pm}$
at the start of the parameter trajectory are depicted as blue and
red dots, and the arrows show the trajectory direction (c) The population
dynamics for clockwise encircling when the system is initialized in
$\protect\ket{\psi_{-}}$. (d) The population dynamics for clockwise
encircling when the system is initialized in $\protect\ket{\psi_{+}}$.
(e) The counterclockwise trajectory (orange) corresponds to the following
parameters in Eq.~(\ref{eq:trajectory_general}): $\delta_{0}=0$,
$g_{0}=0.5$, $R=0.3$, $T=500$, $\omega=2\pi/T$, $\varphi=0$.
(f,g,h) Same as (b,c,d), but for the counterclockwise trajectory.}

\label{fig:chiral-without-encircling}
\end{figure*}

The convetional theoretical \citep{Uzdin2011,Berry2011,Gilary2013,Milburn2015,Hassan2017,Zhong2018,Ribeiro2021}
and experimental \citep{Doppler2016,Pick2019,Choi2020,Liu2020h,Li2022,Chen2021a}
expectation when ecircling an EP is for the state conversion to be
chiral, i.e., dependent on the encircling direction. This is related
to the non-analytic structure of the system eigenvalues $\lambda_{\pm}$
and eigenstates $\ket{\psi_{\pm}}$ at the EP \citep{Zhong2018}.
However, there have also been theoretical \citep{Zhang2019d} and
experimental \citep{Zhang2018} reports about non-chiral behavior
when encircling an EP, depending on the specific trajectory and/or
its starting point.

Consider the trajectory that encircles an EP, as depicted in Figs.~\ref{fig:non-chiral-with-encircling}(a,e).
The naïve theory predicts that for the clockwise encircling, essentially
any initial state is converted into $\ket{\psi_{+}}$ (Figs.~\ref{fig:non-chiral-with-encircling}(c,d)),
whereas for the counterclockwise one, essentially any initial state
is converted into $\ket{\psi_{-}}$ (Figs.~\ref{fig:non-chiral-with-encircling}(g,h)).
Which corresponds to the expected chiral behavior. However, the numerical
curves in these panels tell a different story: for both encircling
directions, numerical simulations predict conversion to $\ket{\psi_{-}}$
--- the non-chiral behavior.

This behavior is easily explainable from our advanced theory. Notice
that for both trajectories the end-point fastest growing state is
$\ket{\psi_{-}}$, see Figs.~\ref{fig:non-chiral-with-encircling}(b,f).
Therefore, the advanced theory predicts non-chiral conversion to $\ket{\psi_{-}}$,
in agreement with the numerical simulations.

We note in passing that apart from numerical and experimental evidence,
Ref.~\citep{Zhang2018} provides analytical considerations to explain
the non-chiral conversion, which --- in contrast to our advanced
theory --- do not include fast noise in the course of evolution.
These considerations turn out to be erroneous. By contrast, Ref.~\citep{Zhang2019d}
provides a correct explanation --- yet one specific to the trajectory
considered in that work; this explanation is consistent with our advanced
theory. We discuss these works in detail in Appendix~\ref{app:Previous_explanations}.

\subsection{\label{subsec:chiral-without-encircling}Chiral state conversion
without encircling the exceptional point}

Another example of unconventional behavior concerns chiral conversion
without encircling any EPs. Such behavior has been reported theoretically
\citep{Hassan2017b,Hassan2017c} and experimentally \citep{Nasari2022}.

Consider the trajectory depicted in Figs.~\ref{fig:chiral-without-encircling}(a,e).
The naïve theory predicts no conversion whatsoever: if the system
is initialized in $\ket{\psi_{-}}$, it ends up in $\ket{\psi_{-}}$
(Figs.~\ref{fig:chiral-without-encircling}(c,g)), and similarly
for $\ket{\psi_{+}}$ (Figs.~\ref{fig:chiral-without-encircling}(d,h)).
This is beacuse this trajectory is fine-tuned: it has $\mathrm{Im}\int_{0}^{T}dt\,\lambda_{\pm}(t)=0$.
However, the numerical simulation predicts conversion to $\ket{\psi_{-}}$
for the clockwise trajectory (Figs.~\ref{fig:chiral-without-encircling}(c,d))
and to $\ket{\psi_{+}}$ for the counterclockwise one (Figs.~\ref{fig:chiral-without-encircling}(g,h)).
Our advanced theory easily explains this: the respective states are
the end-point fastest growing ones, as seen in Figs.~\ref{fig:chiral-without-encircling}(b,f).

Just as in the previous section, we note in passing that works \citep{Hassan2017b,Hassan2017c},
apart from numerical studies, provided analytical arguments to explain
this behavior within the framework of perfectly-controlled slow evolution.
These arguments turn out to be erroneous and are discussed in detail
in Appendix~\ref{app:Previous_explanations}.

\section{Demonstrating the quantitative accuracy of the advanced theory}

\label{sec:quantitative_predictions_advanced_theory}

\begin{figure*}[t]
\begin{centering}
\includegraphics[width=1\textwidth]{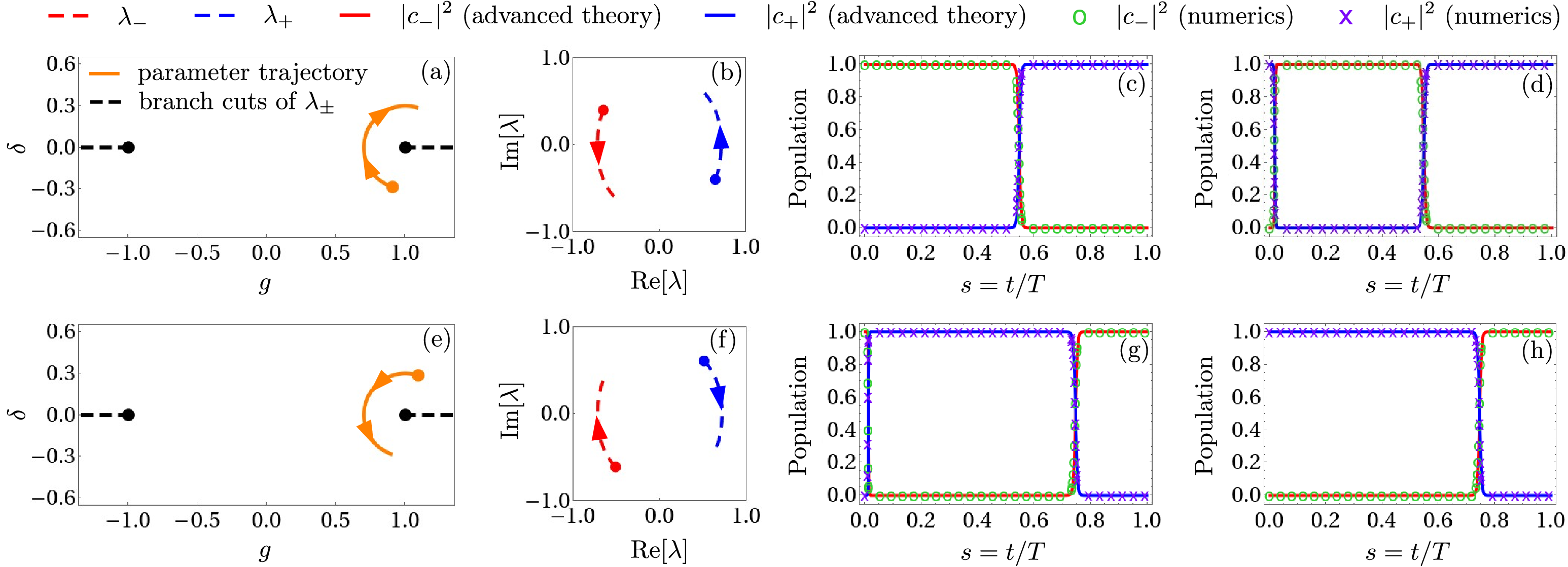}
\par\end{centering}
\caption{\textbf{Open-trajectory evolution under the perturbed Hamiltonian
$\bar{H}(t)$ (\ref{eq:nHH_examples_perturbed}).} The parameter trajectories
in panels (a) and (e) are identical to those in Fig.~\ref{fig:open_trajectory_no_surprises}(a)
and Fig.~\ref{fig:open_trajectory_qualitative_surprise}(a) respectively:
$\delta_{0}=0$, $g_{0}=1$, $R=0.3$, $T=500$, $\omega=-\pi/T$,
$\phi=0.4\pi$ for panel (a) and $\delta_{0}=0$, $g_{0}=1$, $R=0.3$,
$T=500$, $\omega=\pi/T$, $\phi=-0.6\pi$ for panel (e). The trajectory
direction is shown with arrows and the orange dot depict the starting
point. Panels (b) and (f) show the evolution of the instantaneous
eigenvalues of $\bar{H}(t)$. The eigevalues $\lambda_{\pm}$ at the
start of the parameter trajectory are depicted as blue and red dots,
and the arrows show the trajectory direction. Panels (c) and (g) show
the population dynamics along the respective trajectories when starting
in $\protect\ket{\psi_{-}}$, while the panels (d) and (h) show the
same when starting in $\protect\ket{\psi_{+}}$. \textbf{The evolutions
in numerical simulations and as predicted by the advanced theory of
Sec.~\ref{subsec:Summary_Slow-evolution-with-fast-noise} are in
}\textbf{\emph{quantitative}}\textbf{ agreement.}}

\label{fig:open-trajectory-perturbed}
\end{figure*}

In this section, we demonstrate that our advanced theory of Sec.~\ref{subsec:Summary_Slow-evolution-with-fast-noise}
gives not only correct \emph{qualitative} predictions, but is also
capable of accurately predicting the \emph{quantitative} behavior.
We demonstrate this by considering time dynamics that corresponds
to non-Hermitian conversion.

The advanced theory invokes fast perturbation on top of controlled
slow evolution. As we have argued above, such perturbations will appear
generically in experiments and in numerical simulations. However,
in order to demonstrate quantitative accuracy, one needs to know the
dominant source of perturbation. To this end, we consider the evolution
under the Hamiltonian
\begin{equation}
\bar{H}(t)=H(t)+\varepsilon\delta H(t),\label{eq:nHH_examples_perturbed}
\end{equation}
with $H(t)$ as defined in Eqs.~(\ref{eq:nHH_examples}--\ref{eq:trajectory_general})
and
\begin{equation}
\delta H(t)=\begin{pmatrix}0 & \cos\Omega t\\
\cos\Omega t & 0
\end{pmatrix}.\label{eq:deltaH_expression}
\end{equation}
In the examples below, we choose $\varepsilon=10^{-4}$ and $\Omega=2\pi/5$.
The choice of parameters is dictated by the following considerations.
The perturbation size $\varepsilon$ is chosen small in order not
to affect the instantaneous eigenstates and eigenvalues of $\bar{H}$
to any significant degree. The perturbation frequency $\Omega$ is
chosen to be much larger than $\abs{\omega}\leq2\pi/T=2\pi/500$ ($\omega$
being the same as in Figs.~\ref{fig:open_trajectory_no_surprises}--\ref{fig:chiral-without-encircling}).
At the same time, $\Omega$ is chosen to be of the same order as the
distance between the eigenvalues of $H(t)$, $\abs{\lambda_{+}(t)-\lambda_{-}(t)}\sim1$.
This is done in order to ensure that $\varepsilon\delta H(t)$ is
roughly resonant and thus is the dominant perturbation in our numerical
simulations. Other, uncontrolled, perturbations are still inherently
present in our numerical simulations due to time discretization. However,
being close to resonant, the controlled perturbation $\varepsilon\delta H(t)$
is more important than the uncontrolled perturbations.

For our demonstration, we use the same parameter trajectories as considered
in Sec.~\ref{subsec:summary_most_vs_last_growing}. The example of
Fig.~\ref{fig:open_trajectory_no_surprises} considered evolution
along an open trajectory without any controlled perturbations ($\varepsilon=0$).
The naïve theory of Sec.~\ref{subsec:Summary_Perfect-slow-evolution}
was able to predict the outcome of the evolution qualitatively, however,
the time at which the population switch takes place was predicted
incorrectly. The same trajectory is used for Figs.~\ref{fig:open-trajectory-perturbed}(a--d),
now including the controlled perturbation (\ref{eq:nHH_examples_perturbed}--\ref{eq:deltaH_expression}).
By comparing Fig.~\ref{fig:open-trajectory-perturbed}(b) with Fig.~\ref{fig:open_trajectory_no_surprises}(b),
one sees that the perturbation $\varepsilon\delta H(t)$ does not
have any noticeable effect on the Hamiltonian eigenvalues. At the
same time the population switch in the numerical simulation now takes
place at $t_{1}''<0.6T$ (cf.~Figs.\ \ref{fig:open-trajectory-perturbed}(c,d)),
as opposed to $t_{1}'>0.6T$ in Fig.~\ref{fig:open_trajectory_no_surprises}(e).
This shows that the perturbation $\varepsilon\delta H(t)$ of Eq.~(\ref{eq:deltaH_expression})
has a noticeable impact on the evolution, stronger than the numerical
errors. Using the theory of Sec.~\ref{subsec:Summary_Slow-evolution-with-fast-noise},
Eq.~(\ref{eq:summary_evolution_operator_deltaH_perturbation_theory_first_order}),
we calculate the evolution of state populations, as presented in Figs.\ \ref{fig:open-trajectory-perturbed}(c,d).
One sees that the numerical simulation and the predictions of our
advanced theory are in very good quantitative agreement.

The same applies for the reversed trajectory. Figure~\ref{fig:open_trajectory_qualitative_surprise}
showed a population switch at $t_{1}'>0.8T$, which is not predicted
at all by the naïve theory that assumes perfectly-controlled slow
evolution. We explained in Sec.~\ref{subsec:summary_most_vs_last_growing}
how the advanced theory qualitatively predicts this reversal by highlighting
the role of the end-point fastest growing state (as opposed to the
most growing state). Figures~\ref{fig:open-trajectory-perturbed}(e--h)
show that the advanced theory also captures the dynamics quantitatively.
Again, despite no noticeable effect on the eigenvalues in Fig.~\ref{fig:open-trajectory-perturbed}(f)
compared to Fig.~\ref{fig:open_trajectory_qualitative_surprise}(b),
the manually added perturbation $\varepsilon\delta H(t)$ provokes
the population switch at $t_{1}''<0.8T$, cf.~Figs.~\ref{fig:open-trajectory-perturbed}(g,h).
As $t_{1}''<t_{1}'$, this shows the dominance of $\varepsilon\delta H(t)$
over numerical errors. And the result of numerical simulations is
in a very good quantitative agreement with the predictions of the
advanced theory, cf.~Figs.~\ref{fig:open-trajectory-perturbed}(g,h).

We provide details on how the advanced theory curves have been produced
and also further examples of quantitative agreement between the advanced
theory and the numerical simulations in Appendix~\ref{app:quantitative_advanced_theory}.
There we consider closed evolution trajectories associated with chiral
and non-chiral non-Hermitian state conversion.

\section{Proposals for experimental verification}

\label{sec:experimental_verification_proposals}

Below we list a few paradigmatic tests of our theory, that may be
relevant for various experimental platforms.

Recent experiments \citep{Nasari2022} have demonstrated chiral state
transfer when not encircling an EP. Our theory predicts that the conversion
result is determined by the eigenvalues of the non-Hermitian Hamiltonian
towards the end of the evolution. This implies that simply changing
the starting point on the same closed trajectory in the parameter
space can change the conversion behavior from chiral to non-chiral,
cf.~Figs.~\ref{fig:EP-chiral} and \ref{fig:EP-non-chiral} in Appendix~\ref{app:quantitative_advanced_theory}.

More delicate tests of our theory are possible by changing the evolution
speed while still staying within the applicability domain of the theory.
Let us assume that the strength and spectrum of noise do not depend
on details of the controlled evolution. Our theory indicates that
depending on the speed of the controlled evolution during different
segments of the trajectory (e.g., on the evolution time $T$), the
populations of different eigenmodes introduced by the noise will or
will not have enough time to grow, following the analysis of Sec.~\ref{subsec:Summary_Slow-evolution-with-fast-noise}.
When the predictions of our naïve (no noise) and advanced (high-frequency
noise) theories differ, one can switch between the two behaviors by
appropriately choosing $T$ (so that the noise-induced perturbations
do or do not have ample time to be amplified).

Further test of our theory would be to compare runs with the same
initial and final Hamiltonian parameters, but different trajectories
in parameter space. In the appropriate limit (full-fledged noise-induced
perturbations), the final state should not depend on the trajectory.

Finally, in order to verify the quantitative accuracy of our theory,
one could controllably add noise to the system evolution. Making sure
that the controlled noise is dominant over the environmental influences,
one could predict the time dynamics quantitatively, similarly to what
was done in Sec.~\ref{sec:quantitative_predictions_advanced_theory}.

\section{Conclusions}

\label{sec:Conclusions}

Here we have presented a general comprehensive analytic approach for
analyzing slow non-Hermitian dynamics. Our theory comprises two layers.
We first address controllable slow non-Hermitian evolution. We next
add the effect of fast perturbations on top of the slow evolution.
This is a classical way of modeling of high-frequency-noise components
that represent coupling to environment. They are expected both in
experiments and in numerical simulations. Combined with inherent exponential
instabilities of non-Hermitian systems, this second layer of our theory
becomes indispensable when it comes to non-Hermitian dynamics and
radically changes the expected behaviors.

The resulting description greatly simplifies the analysis of such
systems. Simulating the full slow evolution along a certain trajectory
in the Hamiltonian's parameter space is no more needed to infer the
end state. Instead, the resulting state of the system is determined
by the final point in parameter space. It is insensitive to the trajectory
traversed and its time dependence. To experimentally test our theory
one may design different trajectories that end up at the same point
in parameter space and compare the resulting states.

We have applied our theory to analyze the effect of non-Hermitian
state conversion with a special focus on its chirality. We have shown
that our theory allows for predicting the conversion chirality qualitatively,
explaining all the known behaviors. We have further shown that given
the knowledge of the dominant fast perturbation, our theory produces
quantitatively accurate predictions for the specific conversion times.
Ultimately, we have shown that the chirality of slow non-Hermitian
state conversion in the real world is \emph{not} related to the physics
of EPs. Instead, it is determined by the system's decay rate spectrum
at the ends of the evolution trajectory. Hence the chiral state conversion
is non-topological. This provides a simple example of how beautiful
mathematical constructions (the relation between the non-analiticity
at exceptional points and chiral non-Hermitian state conversion) are
washed out by the ``stern realities of life'', i.e., noise.

In practical terms, our theory advances the understanding of slow
non-Hermitian dynamics. The clear intuitive picture it provides can
simplify simulations and design of devices involving few- and many-body,
quantum and classical non-Hermitian systems. Our theory is straightforward
to generalize to more complex descriptions of open systems, such as
quantum systems governed by Lindbladians or hybrid Liouvillians.

\section*{Acknowledgements}

We are grateful to Adi Pick and to Alain Joye for useful discussions.
P.K. acknowledges support from the IIT Jammu Seed Grant (SGT-100106).
Y.G. was supported by the Deutsche Forschungsgemeinschaft (DFG, German
Research Foundation) grant SH 81/8-1 and by a National Science Foundation (NSF-2338819)--Binational Science Foundation (BSF-2023666) grant. K.S. acknowledges
funding by the Deutsche Forschungsgemeinschaft (DFG, German Research
Foundation): under the project number 277101999, CRC 183 (Project
No. C01) and under the project number GO 1405/6-1; by the HQI (\href{http://www.hqi.fr}{www.hqi.fr})
and BACQ initiatives of the France 2030 program financed under the
French National Research Agency grants with numbers ANR-22-PNCQ-0002
and ANR-22-QMET-0002 (MetriQs-France).

The authors have benefited from the hospitality of the International
Centre for Theoretical Sciences (ICTS) (\href{http://www.icts.res.in}{www.icts.res.in})
in Bengaluru, India during the programs ``Condensed matter meets
quantum information'' (2023) and ``Quantum trajectories'' (2025)
(codes: ICTS/COMQUI2023/09 and ICTS/QuTr2025/01), which facilitated
the advancement of this work.

\section*{Code availability}

The Wolfram Mathematica code used to produce the Figures of the manuscript
and the respective simulations can be found on a GitHub repostory
\citep{GitHubRepository}.

\bibliography{bibliography,extras}

\appendix

\clearpage{}

\section{Derivation of analytical results}

\label{app:Derivation_of_analytical_results}

In this Appendix, we provide derivations of our general results for
slow non-Hermitian evolution, which we have concisely presented in
Secs.~\ref{subsec:Summary_Perfect-slow-evolution} and \ref{subsec:Summary_Slow-evolution-with-fast-noise}.
Section~\ref{subsec:Iterative-adiabatic-expansion} focuses on the
case of genuine slow evolution, whereas Sec.~\ref{subsec:Fast-noise+adiabatic-expansion}
considers the effect of uncontrolled fast noise. We emphasize that
the results of this Appendix are valid for arbitrary trajectories
in the parameter space (closed or open) as long, as the evolution
is sufficiently smooth and \emph{no exceptional points or degeneracy
points lie exactly on the trajectory}.

\subsection{\label{subsec:Iterative-adiabatic-expansion}Iterative adiabatic
expansion}

\subsubsection{\label{subsec:adiabatic_expansion_first_order}Idea and first orders}

Consider a time-dependent Hamiltonian $H(t)=H_{s}$ with $s=t/T\in[0,1]$.
The Hamiltonian may be Hermitian or non-Hermitian and is time-dependent.
The values it takes at the initial and final moments of the evolution
are $H(t=0)=H_{0}$ and $H(t=T)=H_{1}$, and the parameter $T\rightarrow\infty$
determines the speed of changing one Hamiltonian into another along
a fixed sequence of Hamiltonians $H_{s}$. We assume that there are
no exceptional or degeneracy points \emph{on} the evolution trajectory,
that is, at all $t\in[0,T]$ the Hamiltonian $H(t)$ can be diagonalised:
\begin{equation}
H(t)\equiv H_{s}=U(t)D(t)U^{-1}(t)\equiv U_{s}D_{s}U_{s}^{-1},\label{eq:H_diagonalization_derivation}
\end{equation}
with $D(t)\equiv D_{s}$ being a diagonal matrix with all diagonal
entries distinct. The diagonal entries of $D(t)$ are the Hamiltonian
eigenvalues $\lambda_{n}(t)\equiv\lambda_{n,s}$. The matrix $U(t)\equiv U_{s}$
encodes the right eigenvectors of the Hamiltonian, $\ket{n(t)}=\ket{n_{s}}=U_{s}\ket n$,
where $\ket n$ is a column vector $\left(0,\ldots,0,1,0,\ldots0\right)^{\mathrm{T}}$
with $1$ on the $n$th postition.

The system state $\ket{\psi(t)}$ obeys the Schrödinger equation
\begin{equation}
i\partial_{t}\ket{\psi(t)}=H(t)\ket{\psi(t)}.
\end{equation}
Our task is to express the final state $\ket{\psi(T)}$ through the
initial state $\ket{\psi(0)}$ in the limit of $T\rightarrow\infty$.
In order to explicitly keep track of the powers of $T$, we switch
to the variable $s$. We define $\ket{\psi(s=t/T)}\equiv\ket{\psi(t)}$.
Note that the quantities with $s$ as a subscript do not depend on
$T$, whereas the evolution of $\ket{\psi(s)}$ does implicitly depend
on $T$ (thence the difference in notation). The Schrödinger equation
now looks
\begin{equation}
i\partial_{s}\ket{\psi(s)}=TH_{s}\ket{\psi(s)}.
\end{equation}

The natural first step for solving this evolution is to work in the
eigenbasis of $H_{s}$. Define $\ket{\varphi(s)}=U_{s}^{-1}\ket{\psi(s)}$.
Then
\begin{equation}
i\partial_{s}\ket{\varphi(s)}=T\left[D_{s}-X^{(1)}(s)\right]\ket{\varphi(s)},
\end{equation}
where $X^{(1)}(s)=\frac{1}{T}iU_{s}^{-1}\partial_{s}U_{s}$. Note
that the term involving $X^{(1)}(s)$ is small compared to the leading
term $D_{s}$. The smallness of $X^{(1)}$, however, does not allow
one to neglect it, as its contribution over the evolution of duration
$T$ may be finite.

In order to arrive at a controllable approximation, introduce $V_{1}(s)$
that diagonalizes $D_{s}-X^{(1)}(s)$:

\begin{equation}
D_{s}-X^{(1)}(s)=V_{1}(s)\Delta_{1}(s)\left[V_{1}(s)\right]^{-1},
\end{equation}
and
\begin{equation}
\ket{\varphi_{1}(s)}=\left[V_{1}(s)\right]^{-1}\ket{\varphi(s)}.
\end{equation}
Here $\Delta_{1}(s)=D_{s}+D^{(1)}(s)$ is diagonal. Within the perturbation
theory in $X^{(1)}$, one finds that
\begin{gather}
V_{1}(s)=\mathbb{I}+W^{(1)}(s),\label{eq:V1}\\
W^{(1)}(s)=\sum_{m\neq n}\ket m\frac{\bra mX^{(1)}(s)\ket n}{\lambda_{m,s}-\lambda_{n,s}}\bra n+O\left(T^{-2}\right)\label{eq:W1_explicit}
\end{gather}
and
\begin{gather}
D^{(1)}(s)=\sum_{n}\ket n\lambda_{n}^{(1)}(s)\bra n,\\
\lambda_{n}^{(1)}(s)=-\bra nX^{(1)}(s)\ket n+O\left(T^{-2}\right).
\end{gather}

Then
\begin{equation}
i\partial_{s}\ket{\varphi_{1}(s)}=T\left[\Delta_{1}(s)-X^{(2)}(s)\right]\ket{\varphi_{1}(s)},\label{eq:phi1_Schr=0000F6dinger}
\end{equation}
where $X^{(2)}(s)=\frac{1}{T}i\left[V_{1}(s)\right]^{-1}\partial_{s}V_{1}(s)$.
Note that $\partial_{s}V_{1}(s)=\partial_{s}W^{(1)}(s)\propto1/T$,
so that $X^{(2)}(s)=O\left(T^{-2}\right)$. Therefore, one can neglect
the contribution of $X^{(2)}(s)$ for the evolution of duration $T$
and approximately write
\begin{equation}
i\partial_{s}\ket{\varphi_{1}(s)}\approx T\Delta_{1}(s)\ket{\varphi_{1}(s)},
\end{equation}
which is immediately solved by
\begin{equation}
\ket{\varphi_{1}(s)}=\exp\left(-iT\int_{0}^{s}d\sigma\Delta_{1}(\sigma)\right)\ket{\varphi_{1}(0)}.
\end{equation}
Note that the time ordering in the exponential is unnecessary since
$\Delta^{(1)}(\sigma)$ is always diagonal.

The solution for $\ket{\varphi(s)}$ follows immediately as
\begin{equation}
\ket{\varphi(s)}=V_{1}(s)\exp\left(-iT\int_{0}^{s}d\sigma\Delta_{1}(\sigma)\right)\left[V_{1}(0)\right]^{-1}\ket{\varphi(0)}.
\end{equation}
Applying this formula for $s=1$ (i.e., $t=T$), one finds
\begin{multline}
\ket{\varphi(s=1)}=\exp\left(-iT\int_{0}^{1}d\sigma\Delta_{1}(\sigma)\right)\ket{\varphi(0)}\\
+W^{(1)}(1)\exp\left(-iT\int_{0}^{1}d\sigma\Delta_{1}(\sigma)\right)\ket{\varphi(0)}\\
+\exp\left(-iT\int_{0}^{1}d\sigma\Delta_{1}(\sigma)\right)W^{(1)}(0)\ket{\varphi(0)}\\
+O\left[T^{-2}\exp\left(-iT\int_{0}^{1}d\sigma\Delta_{1}(\sigma)\right)\right].\label{eq:first_order_adiabatic_solution_full}
\end{multline}
In the exponentials, $T\int_{0}^{1}d\sigma\Delta_{1}(\sigma)=T\int_{0}^{1}d\sigma D(\sigma)+T\int_{0}^{1}d\sigma D^{(1)}(\sigma)$.
$D(s)\propto T^{0}$, so the first integral is proportional to $T$,
whereas the integral of $D^{(1)}(t)$ is proportional to $T^{0}$
and can be neglected.

This leads us to the final formula,
\begin{multline}
\ket{\varphi(s=1)}\approx\exp\left(-iT\int_{0}^{1}d\sigma D(\sigma)\right)\ket{\varphi(0)}\\
+W^{(1)}(1)\exp\left(-iT\int_{0}^{1}d\sigma D(\sigma)\right)\ket{\varphi(0)}\\
+\exp\left(-iT\int_{0}^{1}d\sigma D(\sigma)\right)W^{(1)}(0)\ket{\varphi(0)}.\label{eq:first_order_adiabatic_solution_simplified}
\end{multline}
Let us discuss the meaning of the three terms in this expression.

The first term corresponds to the conventional adiabatic approximation:
each eigenstate of the instantaneous Hamiltonian evolves with the
respective eigenvalue. In the case of a Hermitian $H_{s}$, $D(\sigma)$
is always real, so that each eigenstate simply accumulates the dynamic
phase (and the geometric phase would be given by $T\int_{0}^{1}d\sigma D^{(1)}(\sigma)$
in Eq.~(\ref{eq:first_order_adiabatic_solution_full}), which we
neglected in Eq.~(\ref{eq:first_order_adiabatic_solution_simplified})).
In the non-Hermitian case, the diagonal entries of $D(\sigma)$ may
have an imaginary part, corresponding to some states growing and some
decaying. Therefore, for sufficiently large $T$, one expects this
term to align with the most growing (least decaying) eigenstate, i.e.,
the state $\ket{n_{s=1}}$, for which the integral of the growth rate
over the entire trajectory, $\mathrm{\mathrm{Im}\,}\int_{0}^{1}ds\lambda_{n,s}$,
is the largest. If, however, the most growing state has zero amplitude
in the intial state $\ket{\varphi(0)}$, then this term aligns with
the second most growing state (unless the amplitude of this state
is also zero, and so on).

The second term in Eq.~(\ref{eq:first_order_adiabatic_solution_simplified})
tells that due to finite $T$, the adiabatic limit is violated, and
the final state $\ket{\varphi(s=1)}$ is only equal to the most growing
eigenstate up to $O(T^{-1})$ corrections. Finally, the last term
implies that due to adiabaticity violation at the beginning of the
evolution, population of all $\ket{n_{s=0}}$ is generically generated,
and therefore the above remark concerning the amplitudes in the initial
state is insignificant: generically, for sufficiently large $T$,
the state $\ket{\varphi(s=1)}$ aligns with the most growing among
the eigenstates $\ket{n_{s=1}}$.

Equation (\ref{eq:first_order_adiabatic_solution_simplified}) thus
predicts and describes the non-Hermitian state conversion: essentially
any initial state under slow non-Hermitian evolution gets converted
into the most growing eigenstate. All the other states are suppressed.
Naïvely, one expects this suppression to be exponential in $T$. However,
the conversion accuracy is only $O(T^{-1})$ due to adiabaticity violation
at the end of the evolution.\footnote{Note that the correction in the exponential $T\int_{0}^{1}d\sigma D^{(1)}(\sigma)=O(1)$
is insignificant in the limit $T\rightarrow\infty$ compared to the
main term $O(T)$ and cannot change which eigenstate $\ket{n_{s=1}}$
is preferred. Unless there is a degeneracy of $\int_{0}^{1}d\sigma D(\sigma)$,
in which case the degeneracy breaking does not scale with $T\rightarrow\infty$
and thus still cannot result in perfect conversion.

Some effects of the non-Hermitian Berry phase for finite $T$ have
been recently discussed in Ref.~\citep{Longhi2023}.}

\subsubsection{Solution to an arbitrary order}

Having demonstrated in Sec.~\ref{subsec:adiabatic_expansion_first_order}
that the lowest order of the adiabatic expansion captures the physics
of the non-Hermitian state conversion, we are now in a position to
develop a systematic expansion in powers of $T$ and demonstrate that
further orders do not bring qualitative changes. Instead of neglecting
$X^{(2)}$ in Eq.~(\ref{eq:phi1_Schr=0000F6dinger}), define $V_{2}(s)$
that diagonalizes $\Delta_{1}(s)-X^{(2)}(s)$:

\begin{equation}
\Delta_{1}(s)-X^{(2)}(s)=V_{2}(s)\Delta_{2}(s)\left[V_{2}(s)\right]^{-1},
\end{equation}
and
\begin{equation}
\ket{\varphi_{2}(s)}=\left[V_{2}(s)\right]^{-1}\ket{\varphi_{1}(s)}.
\end{equation}
Here $\Delta_{2}(s)=\Delta_{1}(s)+D^{(2)}(s)$ is diagonal. Within
the perturbation theory in $X^{(2)}$, one finds that
\begin{gather}
V_{2}(s)=\mathbb{I}+W^{(2)}(s),\label{eq:V2}\\
W^{(2)}(s)=\sum_{m\neq n}\ket m\frac{\bra mX^{(2)}(s)\ket n}{\lambda_{m,s}-\lambda_{n,s}}\bra n+O\left(T^{-3}\right)
\end{gather}
and
\begin{gather}
D^{(2)}(s)=\sum_{n}\ket n\lambda_{n}^{(2)}(s)\bra n,\\
\lambda_{n}^{(2)}(s)=-\bra nX^{(2)}(s)\ket n+O\left(T^{-3}\right).
\end{gather}
Then
\begin{equation}
i\partial_{s}\ket{\varphi_{2}(s)}=T\left[\Delta_{2}(s)-X^{(3)}(s)\right]\ket{\varphi_{2}(s)},\label{eq:phi2_Schr=0000F6dinger}
\end{equation}
where $X^{(3)}(s)=\frac{1}{T}i\left[V_{2}(s)\right]^{-1}\partial_{s}V_{2}(s)$.
Note that $\partial_{s}V^{(1)}(s)=\partial_{s}W^{(2)}(s)\propto1/T^{2}$,
so that $X^{(3)}(s)=O\left(T^{-3}\right)$.

Similarly, one can iteratively define for any order $k$
\begin{multline}
\ket{\varphi_{k}(s)}=\left[V_{k}(s)\right]^{-1}\ket{\varphi_{k-1}(s)}\\
=\left[V_{k}(s)\right]^{-1}\ldots\left[V_{1}(s)\right]^{-1}\ket{\varphi(s)},
\end{multline}
obeying
\begin{equation}
i\partial_{s}\ket{\varphi_{k}(s)}=T\left[\Delta_{k}(s)-X^{(k)}(s)\right]\ket{\varphi_{k}(s)},\label{eq:phi_k_Schr=0000F6dinger}
\end{equation}
with
\begin{equation}
X^{(k)}(s)=\frac{1}{T}i\left[V_{k-1}(s)\right]^{-1}\partial_{s}V_{k-1}(s)=O\left(T^{-k}\right).
\end{equation}
The matrices $V_{k}(s)$ are defined as diagonalisers of $\Delta_{k-1}(s)-X^{(k-1)}(s)$:
\begin{equation}
\Delta_{k-1}(s)-X^{(k-1)}(s)=V_{k}(s)\Delta_{k}(s)\left[V_{k}(s)\right]^{-1},
\end{equation}
\begin{equation}
\Delta_{k}(s)=\Delta_{k-1}(s)+D^{(k)}(s),\quad D^{(k)}=O\left(T^{-k}\right),
\end{equation}
\begin{equation}
V_{k}(s)=\mathbb{I}+W^{(k)}(s),\quad W^{(k)}(s)=O\left(T^{-k}\right).\label{eq:Vk}
\end{equation}
Note that we mark the objects of order $T^{-k}$ by the superscript
$(k)$, such as $X^{(k)}$ or $W^{(k)}$, whereas other objects appearing
at $k$th iteration and having subscript $k$, like $V_{k}$, are
not small.

Neglecting the small term $X^{(k)}(s)\sim T^{-k}$ in Eq.~(\ref{eq:phi_k_Schr=0000F6dinger}),
one then obtains that
\begin{equation}
\ket{\varphi(s)}\approx U_{k}(s)\exp\left(-iT\int_{0}^{s}d\sigma\Delta_{k}(\sigma)\right)\left[U_{k}(0)\right]^{-1}\ket{\varphi(0)},
\end{equation}
\begin{equation}
U_{k}(s)=V_{1}(s)V_{2}(s)\ldots V_{k}(s)=\mathbb{I}+W_{k}(s),\label{eq:W_k}
\end{equation}

\begin{equation}
\left[U_{k}(s)\right]^{-1}=\left[V_{k}(s)\right]^{-1}\ldots\left[V_{2}(s)\right]^{-1}\left[V_{1}(s)\right]^{-1}=\mathbb{I}+\tilde{W}_{k}(s),\label{eq:Wtilde_k}
\end{equation}
Therefore,
\begin{multline}
\ket{\varphi(s=1)}=\exp\left(-iT\int_{0}^{1}d\sigma\Delta_{k}(\sigma)\right)\ket{\varphi(0)}\\
+W_{k}(1)\exp\left(-iT\int_{0}^{1}d\sigma\Delta_{k}(\sigma)\right)\ket{\varphi(0)}\\
+\exp\left(-iT\int_{0}^{1}d\sigma\Delta_{k}(\sigma)\right)\tilde{W}_{k}(0)\ket{\varphi(0)}\\
+W_{k}(1)\exp\left(-iT\int_{0}^{1}d\sigma\Delta_{k}(\sigma)\right)\tilde{W}_{k}(0)\ket{\varphi(0)}\\
+O\left(T^{-k-1}\exp\left(-iT\int_{0}^{1}d\sigma\Delta_{k}(\sigma)\right)\right).\label{eq:kth_order_adiabatic_solution_full}
\end{multline}
Notice that $\Delta_{k}(s)=D(s)+O\left(T^{-1}\right)$, so the physics
described by this formula is the same as the one of Eq.~(\ref{eq:first_order_adiabatic_solution_simplified}).
The only significant difference is the appearance of the fourth term,
involving both $W_{k}$ and $\tilde{W}_{k}$, each of which is $O(T^{-1})$.
This term shows that the non-Hermitian conversion is never exponentially
perfect but is only accurate up to $O(T^{-1})$ corrections due to
non-adiabaticity at the end of the trajectory --- something hinted
at already by the second term in Eq.~(\ref{eq:first_order_adiabatic_solution_simplified}).

\subsubsection{Convergence and errors}

Remark that the solution (\ref{eq:kth_order_adiabatic_solution_full})
is valid to order $T^{-k}$ for arbitrary $k$, and is therefore asymptotically
exact. It is important to note, however, that the iterations do not
necessarily \emph{converge} to the exact solution as $k\rightarrow\infty$.
I.e., our expansion may be an asymptotic expansion, but not a convergent
one.

In order to appreciate the difference, consider the Landau-Zener problem
for \emph{Hermitian} Hamiltonians \citep{Zener1932,Landau1932a,Stueckelberg1932,Majorana1932,Landau1977}.
Consider a two-level Hamiltonian
\begin{equation}
H(t)=\begin{pmatrix}\frac{1}{2}\lambda t & \eta\\
\eta & -\frac{1}{2}\lambda t
\end{pmatrix}\label{eq:LZ_Hamiltonian}
\end{equation}
controlling the system evolution from $t=-t_{0}/2$ until $t=t_{0}/2$;
we take $\lambda>0$ and $\eta>0$ for simplicity. For the evolution
time $t_{0}\rightarrow\infty$ the initial state $\left(1,0\right)^{\mathrm{T}}$
(which is the ground state at $t\rightarrow-\infty$) may become at
$t\rightarrow\infty$ the new ground state $\left(0,1\right)^{\mathrm{T}}$
or the excited state $\left(1,0\right)^{\mathrm{T}}$. The latter
happens with the Landau-Zener probability $P_{\mathrm{LZ}}=\exp\left(-2\pi\eta^{2}/\lambda\right)$
\citep{Zener1932,Landau1932a,Stueckelberg1932,Majorana1932,Landau1977}.
The speed of evolution is controlled by $\lambda$. In particular,
for $\lambda\rightarrow0$, $P_{\mathrm{LZ}}$ vanishes, in agreement
with the adiabatic theorem. For small but finite $\lambda$, $P_{\mathrm{LZ}}$
is exponentially small and is non-analytic as $\lambda\rightarrow0$.

Now compare this to the predictions of Eq.~(\ref{eq:kth_order_adiabatic_solution_full}).
In order to make a link with our definitions, introduce $E_{0}=\lambda t_{0}$,
$T=t_{0}$, and take the limit of $t_{0}\rightarrow\infty$ such that
$E_{0}=\mathrm{const}$. That is, $\lambda=E_{0}/t_{0}\rightarrow0$
and $T=E_{0}/\lambda\rightarrow\infty$. The matrix diagonalizing
the Hamiltonian (\ref{eq:LZ_Hamiltonian}) is $U(t)\equiv U_{s}=\mathbb{I}+O(\eta/E_{0})$,
cf.~Eq.~(\ref{eq:H_diagonalization_derivation}). Therefore, $W_{k}(1)$
and $\tilde{W}_{k}(0)$ in Eq.~(\ref{eq:kth_order_adiabatic_solution_full})
are $\propto\eta/E_{0}$, so that in the Landau-Zener limit $E_{0}\rightarrow\infty$,
the equation predicts vanishing probability up to corrections of order
$T^{-k-1}$ for any $k$. On one hand, this is \emph{consistent} with
the exact Landau-Zener result $P_{\mathrm{LZ}}=\exp\left(-.../\lambda\right)=\exp\left(-...T\right)$.
On the other hand, expansion~(\ref{eq:kth_order_adiabatic_solution_full})
does not enable one to obtain the exponentially small answer.

Two remarks are due in relation to the above comparison. First, we
have deliberately avoided a discussion of the order of limits above.
However, a proper rewriting of the Landau-Zener formula in the newly-defined
units, $P_{\mathrm{LZ}}=\exp\left(-2\pi\eta^{2}T/E_{0}\right)$, hints
that the exact way of taking limits $\eta T\rightarrow\infty$ and
$E_{0}/\eta\rightarrow\infty$ is important. The ways they are taken
in the Landau-Zener problem ($E_{0}\rightarrow\infty$ while keeping
$E_{0}/T$ constant) and in the derivation of expansion~(\ref{eq:kth_order_adiabatic_solution_full})
(first $T\rightarrow\infty$, and then $E_{0}\rightarrow\infty$)
are different. Therefore, the above comparison to the Landau-Zener
problem only serves to illustrate the possibility of effects not captured
by our expansion, but not to prove their existence.

Second, when considering finite $E_{0}$ (equivalently, finite $t_{0}$),
the Landau-Zener formula aqcuires corrections, which are power-law
in $T$ \citep{Rubbmark1981,Mullen1989,Lubin1990,Shimshoni1991,Shimshoni1993}.
We expect our expansion to work in this setting. However, we have
not investigated this problem in detail.

We note in passing that a number of works have investigated non-Hermitian
Landau-Zener problems \citep{Kvitsinsky1991a,Malla2023}. Further,
some recent works considered Hermitian Landau-Zener problems in the
presence of Lindbladian terms \citep{Avron2011a,Fraas2017a,Joye2022a}.

\subsubsection{Subtleties}

Here we emphasize a few subtleties which do not influence the validity
of Eq.~(\ref{eq:kth_order_adiabatic_solution_full}), yet should
be taken into account when interpreting its implications.

\emph{Corrections to non-Hermitian state conversion}. The first and
third terms in Eq.~(\ref{eq:kth_order_adiabatic_solution_full})
describe non-Hermitian state conversion. Non-adiabaticity at the beginning
of the evolution leads to generating amplitudes on all the instantaneous
eigenstates. Then due to the eigenvalues having an imaginary part,
the final state aligns with the most growing eigenstate $\ket{n_{s}}$
as $T\rightarrow\infty$. Based on these two terms, one may conclude
that the conversion is exponentially accurate, and the contributions
of other eigenstates are suppressed by factors of $\exp(-\ldots T)$.
The second and fourth terms, however, clearly show that the non-adiabaticity
at the end of the evolution generates transitions from the most growing
eigenstate $\ket{n_{s}}$ to other eigenstates $\ket{m_{s}}$. Given
that $W_{k}$ is generically $\propto T^{-1}$, this means that the
conversion is only power-law accurate. This has been previously pointed
out in Ref.~\citep{Ribeiro2021}.

\emph{Non-generic Hamiltonians}. The above interpretation of conversion
is based on the assumtion that the matrix $\tilde{W}_{k}(0)$ has
all off-diagonal matrix elements non-zero, i.e., the non-adiabaticity
at the beginning of the evolution generates components on all instantaneous
eigenstates of the Hamiltonian. While generically true, this may not
be the case for Hamiltonians featuring a special structure. For example,
if $H_{s}$ admits a symmetry $\mathcal{S}$, $\left[H_{s},\mathcal{S}\right]=0$
for all ``times'' $s$, then all the matrix elements of $\tilde{W}_{k}(0)$
connecting states in different symmetry sectors will vanish. I.e.,
the conversion will happen to the most growing state within the symmetry
sector. Similarly, if the Hamiltonian concerns two distinct subsystems
(i.e., $H_{s}=H_{1,s}\otimes\mathbb{I}+\mathbb{I}\otimes H_{2,s}$)
and the evolution concerns only one of them ($H_{2,s}\equiv H_{2,0}$),
the conversion will only concern the part of the state related to
the first system, while the second component of the state will remain
intact.

\subsection{\label{subsec:Fast-noise+adiabatic-expansion}The effect of fast
noise}

The consideration of Sec.~\ref{subsec:Iterative-adiabatic-expansion}
assumed a controlled adiabatic evolution. However, in practice the
evolution is never perfectly controlled due to systematic deviations
and noise in the system parameters. Non-Hermitian systems feature
exponential instabilities, making it mandatory to investigate the
effect of such imperfections. Systematic imperfections and noise that
is adiabatically slow, do not generate qualitatively new effects.
They can be incorporated into the theory of Sec.~\ref{subsec:Iterative-adiabatic-expansion}
as part of the system Hamiltonian. At most, they may change the specific
form of the eigenstates and which of the eigenstates grows the most
over the course of the evolution.

In contrast non-adiabatic noise may lead to qualitatively new effects
by generating non-adiabatic transitions amidst the evolution. Below
we show that this leads to a qualitative change: it is not the most
growing state that wins in the course of the evolution, but the state
that grows the last along the trajectory (cf.~Sec.~\ref{subsec:Summary_Slow-evolution-with-fast-noise}).
Below we derive the theory of non-Hermitian state conversion in the
presence of fast (non-adiabatic) noise and explain its implications.

\subsubsection{Derivation}

In Sec.~\ref{subsec:Iterative-adiabatic-expansion}, we have shown
that the evolution under time-dependent Hamiltonian $H(t)=H_{s}$
with $s=t/T\in[0,1]$ for $T\rightarrow\infty$ is governed by the
operator
\begin{multline}
\mathcal{U}(T,0)=\mathcal{T}\exp\left(-i\int_{0}^{T}dt\,H(t)\right)\\
=U(T)\mathcal{E}(T,0)U^{-1}(0),\label{eq:kth_order_adiabatic_solution_full_operator_form_1}
\end{multline}
\begin{multline}
\mathcal{E}(T,0)=\exp\left(-iT\int_{0}^{1}d\sigma\Delta_{k}(\sigma)\right)\\
+W_{k}(1)\exp\left(-iT\int_{0}^{1}d\sigma\Delta_{k}(\sigma)\right)\\
+\exp\left(-iT\int_{0}^{1}d\sigma\Delta_{k}(\sigma)\right)\tilde{W}_{k}(0)\\
+W_{k}(1)\exp\left(-iT\int_{0}^{1}d\sigma\Delta_{k}(\sigma)\right)\tilde{W}_{k}(0)\ket{\varphi(0)}\\
+O\left(T^{-k-1}\exp\left(-iT\int_{0}^{1}d\sigma\Delta_{k}(\sigma)\right)\right),\label{eq:kth_order_adiabatic_solution_full_operator_form_2}
\end{multline}
where $U(t)$ encodes the instantaneous eigenbasis of $H(t)$, cf.~Eq.~(\ref{eq:H_diagonalization_derivation}),
$\Delta_{k}(s)=D_{s}+O\left(T^{-1}\right)$ are diagonal matrices
$D_{s}$ encoding the instantaneous eigenvalues $\lambda_{n}(t)=\lambda_{n,s}$
of $H(t)$ plus some unimportant non-adiabatic corrections, $W_{k}(s)$
and $\tilde{W}_{k}(s)$ encode the non-adiabatic effects of eigenbasis
evolution along the trajectory, and the form of $\mathcal{E}(T,0)$
is readily inferred from Eq.~(\ref{eq:kth_order_adiabatic_solution_full}).
The operator $\mathcal{U}(T,0)$ predicts conversion of essentially
any initial state into one specific eigenstate $\ket{n(T)}$ of $H(T)$
in the limit $T\rightarrow\infty$; the final eigenstate $\ket n$
is the state that has the largest integral $\mathrm{Im}\int_{0}^{1}d\sigma\lambda_{n,\sigma}$
--- that is, the eigenstate that grows the most over the course of
the evolution.

Consider now the evolution generated by $\bar{H}(t)=H(t)+\varepsilon\delta H(t)$,
where $\delta H(t)$ is a correction to the Hamiltonian due to imperfect
control of the system parameters and $\varepsilon\ll1$ controls its
smallness. If this correction can be incorporated in the framework
of slow evolution, such that one can write $\bar{H}(t=sT)=\bar{H}_{s}$,
the predictions of Eqs.~(\ref{eq:full_adiabatic_solution_arbitrary_times_1}--\ref{eq:full_adiabatic_solution_arbitrary_times_2})
hold: there is conversion to the most growing eigenstate of $\bar{H}$.
The adiabatic perturbation may change the exact form of the eigenstate
or may change which of the eigenstates wins. In contrast, we are interested
in the effect of fast perturbations, which cannot be incorporated
into the framework of iterative adiabatic expansion of Sec.~\ref{subsec:Iterative-adiabatic-expansion}.

We are interested in calculating the evolution operator
\begin{equation}
\mathcal{\bar{U}}(T,0)=\mathcal{T}\exp\left(-i\int_{0}^{T}dt\,\bar{H}(t)\right),
\end{equation}
which under the perturbation theory in $\delta H(t)$ can be expressed
as
\begin{multline}
\mathcal{\bar{U}}(T,0)=\mathcal{U}(T,0)\\
-i\varepsilon\int_{0}^{T}dt_{1}\mathcal{U}(T,t_{1})\delta H(t_{1})\mathcal{U}(t_{1},0)\\
-\varepsilon^{2}\int_{0}^{T}dt_{2}\int_{0}^{t_{2}}dt_{1}\mathcal{U}(T,t_{2})\delta H(t_{2})\mathcal{U}(t_{2},t_{1})\delta H(t_{1})\mathcal{U}(t_{1},0)\\
+O\left(\varepsilon^{3}\right),\label{eq:evolution_operator_deltaH_perturbation_theory}
\end{multline}
where
\begin{equation}
\mathcal{U}(t_{2},t_{1})=\mathcal{T}\exp\left(-i\int_{t_{1}}^{t_{2}}dt\,H(t)\right).
\end{equation}

Applying the result of Sec.~\ref{subsec:Iterative-adiabatic-expansion}
to the evolution between $t_{1}=s_{1}T$ and $t_{2}=s_{2}T$, one
can write that
\begin{equation}
\mathcal{U}(t_{2},t_{1})=U(t_{2})\mathcal{E}(t_{2},t_{1})U^{-1}(t_{1}),\label{eq:full_adiabatic_solution_arbitrary_times_1}
\end{equation}
\begin{multline}
\mathcal{E}(t_{2},t_{1})=\exp\left(-iT\int_{s_{1}}^{s_{2}}d\sigma\Delta_{k}(\sigma)\right)\\
+W_{k}(s_{2})\exp\left(-iT\int_{s_{1}}^{s_{2}}d\sigma\Delta_{k}(\sigma)\right)\\
+\exp\left(-iT\int_{s_{1}}^{s_{2}}d\sigma\Delta_{k}(\sigma)\right)\tilde{W}_{k}(s_{1})\\
+W_{k}(s_{2})\exp\left(-iT\int_{s_{1}}^{s_{2}}d\sigma\Delta_{k}(\sigma)\right)\tilde{W}_{k}(s_{1})\ket{\varphi(0)}\\
+O\left(T^{-k-1}\exp\left(-iT\int_{s_{1}}^{s_{2}}d\sigma\Delta_{k}(\sigma)\right)\right).\label{eq:full_adiabatic_solution_arbitrary_times_2}
\end{multline}
In other words, for slow evolution between $t_{1}$ and $t_{2}$,
essentially any initial state at $t_{1}$ becomes one specific eigenstate
$\ket{n(t_{2})}$ at the end of the time interval. The specific eigenstate
is determined by the largest integral $\mathrm{Im}\int_{s_{1}}^{s_{2}}d\sigma\lambda_{n,\sigma}$
--- i.e., the eignestate that grows the most during this interval
wins.

The implication of the above conclusion for the perturbed evolution
is drastic. To the first order in $\delta H$,
\begin{multline}
\mathcal{\bar{U}}(T,0)=\mathcal{U}(T,0)\\
-i\varepsilon\int_{0}^{T}dt_{1}\mathcal{U}(T,t_{1})\delta H(t_{1})\mathcal{U}(t_{1},0)\\
+O\left(\varepsilon^{2}\right).\label{eq:evolution_operator_deltaH_perturbation_theory_first_order}
\end{multline}
Note that $\mathcal{U}(T,0)=\mathcal{U}(T,t_{1})\mathcal{U}(t_{1},0)$
for any $t_{1}$, so that the role of $\delta H(t)$ is to disrupt
the unperturbed (noiseless) evolution at all possible intermediate
times. Focus on $\varepsilon\delta H(t_{1})\mathcal{U}(t_{1},0)$.
By the time $t_{1}=s_{1}T$, essentially any initial state becomes
aligned with one specific eigenstate $\ket{n(t_{1})}$ due to the
unperturbed evolution $\mathcal{U}(t_{1},0)$. The perturbation at
time $t_{1}$ generates amplitude of order $\varepsilon$ in other
(generically --- all other) eigenstates of $H(t_{1})$. This effectively
restarts the non-Hermitian state conversion process. If the preferred
state of $\mathcal{U}(T,t_{1})$ for some $t_{1}$ is different from
that of $\mathcal{U}(T,0)$, the evolution outcome may change. We
discuss the details of this in Sec.~\ref{subsec:perturbation_physical_implications}.

\subsubsection{\label{subsec:perturbation_physical_implications}Physical meaning
and implications}

In order to understand, how the perturbation term in Eq.~(\ref{eq:evolution_operator_deltaH_perturbation_theory_first_order})
can change the outcome of the evolution, consider the following simplified
example. Consider a two-level system with $\mathrm{Im}\,\lambda_{1,s}>\mathrm{Im}\,\lambda_{2,s}$
for $s\in[0,s_{0})$, $\mathrm{Im}\,\lambda_{1,s}<\mathrm{Im}\,\lambda_{2,s}$
for $s\in(s_{0},1]$, where $0<s_{0}<1$, and $\mathrm{Im}\int_{0}^{1}d\sigma\lambda_{1,\sigma}>\mathrm{Im}\int_{0}^{1}d\sigma\lambda_{2,\sigma}$.\footnote{Sec.~\ref{subsec:summary_most_vs_last_growing} and Fig.~\ref{fig:open_trajectory_qualitative_surprise}
in it provide such an example up to the identification $n=1=+$, $n=2=-$.} The last statement implies that the unperturbed (noiseless) evolution
$\mathcal{U}(T,0)$ converts essentially any state into the first
eigenstate, $\ket{n=1}$. However, the late parts of the evolution,
$\mathcal{U}(T,sT)$ for any $s\geq s_{0}$, have a different preference:
$\mathrm{Im}\int_{s\geq s_{0}}^{1}d\sigma\lambda_{1,\sigma}<\mathrm{Im}\int_{s\geq s_{0}}^{1}d\sigma\lambda_{2,\sigma}$.
The late parts of the evolution would convert essentially any initial
state to $\ket{n=2}$. The reason why this does not happen in the
unperturbed evolution is because by $t=s_{0}T$, the system is aligned
with $\ket{n=1}$ up to corrections $\sim\exp\left(T\mathrm{Im}\int_{0}^{s_{0}}d\sigma\left[\lambda_{2,\sigma}-\lambda_{1,\sigma}\right]\right)$;
the exponential growth of $\ket{n=2}$ amplitude relative to that
of $\ket{n=1}$ for $t>s_{0}T$, $\exp\left(T\mathrm{Im}\int_{s_{0}}^{1}d\sigma\left[\lambda_{2,\sigma}-\lambda_{1,\sigma}\right]\right)$,
is smaller than its suppression in the first part of the evolution:
$\exp\left(T\mathrm{Im}\int_{0}^{1}d\sigma\left[\lambda_{2,\sigma}-\lambda_{1,\sigma}\right]\right)\ll1$.

In the presence of the perturbation, the situation changes drastically.
The dominant eigenstate now dominates by the factor of $\varepsilon^{-1}$
only --- not exponentially in $T$. For example, if the state at
$s=s_{0}$ is $\ket{n=1}$, the perturbation at $s_{0}$ will create
amplitude $O(\varepsilon)\gg\exp\left(-T\mathrm{Im}\int_{0}^{s_{0}}d\sigma\left[\lambda_{2,\sigma}-\lambda_{1,\sigma}\right]\right)$
for $\ket{n=2}$. Then the final state at $t=T$ will be\begin{widetext}
\begin{equation}
\underbrace{\exp\left(T\mathrm{Im}\int_{s_{0}}^{1}d\sigma\lambda_{1,\sigma}\right)\ket{n=1}}_{\mathcal{U}(T,s_{0}T)\ket{n=1}}+\underbrace{\varepsilon\exp\left(T\mathrm{Im}\int_{s_{0}}^{1}d\sigma\lambda_{2,\sigma}\right)\ket{n=2}}_{\varepsilon\mathcal{U}(T,s_{0}T)\delta H(s_{0}T)\ket{n=1}}.
\end{equation}
\end{widetext}The ratio of the second and the first amplitudes is
$\varepsilon\exp\left(T\mathrm{Im}\int_{s_{0}}^{1}d\sigma\left[\lambda_{2,\sigma}-\lambda_{1,\sigma}\right]\right)\gg1$
for $T\rightarrow\infty$. Therefore, the perturbation at $t=s_{0}T$
(or any later time) makes the system align with $\ket{n=2}$ state.

This example illustrates that the presence of fast perturbations drastically
changes the properties of non-Hermitian state conversion during slow
evolution: while Eqs.~(\ref{eq:full_adiabatic_solution_arbitrary_times_1}--\ref{eq:full_adiabatic_solution_arbitrary_times_2})
predict conversion of essentially any initial state to the eigenstate
that grows the most in the course of the evolution duration, Eq.~(\ref{eq:evolution_operator_deltaH_perturbation_theory_first_order})
predicts conversion to the eigenstate that grows the most \emph{in
the latest stint} of the evolution. The latter is the fundamental
reason why the chirality of non-Hermitian state conversion is not
tied to encircling exceptional points, as illustrated in the main
text.

\section{\label{app:Previous_explanations}Discussion of previous explanations
of unusual non-Hermitian state conversion and their connection to
our advanced theory of Sec.~\ref{subsec:Summary_Slow-evolution-with-fast-noise}}

We have demonstrated in Sec.~\ref{subsec:Summary_Implications} that
the theory introduced in Sec.~\ref{subsec:Summary_Slow-evolution-with-fast-noise}
predicts and explains the phenomenology of non-Hermitian state conversion,
including the exceptions to the naïvely expected behavior. The latter
are puzzling to understand without invoking uncontrolled fast perturbations.
At the same time, these exceptions have been previously discussed
in the literature and have been provided with some theoretical explanations
--- without talking of uncontrolled perturbations! Here we discuss
the instances of such explanations known to us. We show that two out
of three explanations have errors in them. One explanation is correct,
yet specific for a class of trajectories considered in that work;
this one is consistent with our general theory of Sec.~\ref{subsec:Summary_Slow-evolution-with-fast-noise}.
We thus claim that the advanced analytical theory introduced in the
present work is comprehensive in describing non-Hermitian state conversion
under slow evolution.

\subsection{Non-chiral dynamics with EP encircling}

\uline{Reference \mbox{\citep{Zhang2018}}} has reported experimental,
numerical, and analytical results demonstrating non-chiral conversion
dynamics when encircling an EP. We have no reasons to doubt the experimental
and numerical results presented in Ref.~\citep{Zhang2018}. However,
we will now show that the analytical considerations presented in that
paper in Section ``V. Theoretical demonstration of the nonchiral
dynamics'' contain an interpretation error and do not actually explain
non-chiral dynamics upon encircling a single~EP.

In the section in question, Ref.~\citep{Zhang2018} considers the
Hamiltonian identical to our Eq.~(\ref{eq:nHH_examples}) and trajectories
that can be described by our Eq.~(\ref{eq:trajectory_general}) with
$\delta_{0}=0$, $g_{0}=1$, and $\phi=\pi/2$. That is, the same
trajectory as in our Sec.~\ref{subsec:non-chiral-with-encircling}
modulo a change in the starting point. Section V of Ref.~\citep{Zhang2018}
proceeds to solve the Schrödinger equation, $i\partial_{t}\ket{\psi}=H(t)\ket{\psi}$.
The authors find an exact solution in terms of hypergeometric functions
and demonstrate non-chiral behavior in the limit of the trajectory
radius $R\rightarrow\infty$. Evidently, for $R\rightarrow\infty$,
the trajectory encircles both EPs present in the system. Indeed, in
this case no chiral dynamics is expected for perfectly-controlled
slow evolution, cf.~Ref.~\citep{Zhong2018}, --- in agreement with
the above result of Ref.~\citep{Zhang2018}. However, clearly, this
calculation does not explain the absence of chiral dynamics upon encircling
a single EP.

\uline{Reference \mbox{\citep{Zhang2019d}}} has reported numerical
and analytical results demonstrating non-chiral conversion dynamics
when encircling an EP. We believe their results to be correct and
will now discuss their analytical part, presented in Section ``V.
Theoretical demonstration of the dynamics'' of Ref.~\citep{Zhang2019d}.

Reference \citep{Zhang2019d} also employs the Hamiltonian identical
to our Eq.~(\ref{eq:nHH_examples}). However, the trajectory studied
in that paper is highly non-trivial. The trajectory is continuous,
but not smooth. It consists of three smooth pieces, see Figs.~2 and
8 of Ref.~\citep{Zhang2019d}. The authors find an exact solution
for the last part of the trajectory (again, in terms of hypergeometric
functions). They show that the prediction of conversion during the
last part of the trajectory coincides with the results of numerical
simulations and explains the non-chiral dynamics.

Notice that this explanation is strikingly similar to the prediction
of our advanced theory --- the last part of the trajectory determines
the conversion outcome. At the same time, Ref.~\citep{Zhang2019d}
does not consider fast noise, which is an essential ingredient of
our advanced theory. Why does the explanation of Ref.~\citep{Zhang2019d}
work then? The reason is the non-smooth trajectory they consider.

The predictions of our naïve theory in Eqs.~(\ref{eq:summary_perfectly_slow_evolution_1}--\ref{eq:summary_perfectly_slow_evolution_2})
are valid for infinitely smooth trajectories. For a non-smooth (yet
perfectly controlled and slow) trajectory, Eqs.~(\ref{eq:summary_perfectly_slow_evolution_1}--\ref{eq:summary_perfectly_slow_evolution_2})
can be applied to each smooth part, and then the evolution operators
for each part should be multiplied in order to obtain the evolution
operator for the whole trajectory. Upon the multiplication, the operators
$W$ and $\tilde{W}$ in Eq.~(\ref{eq:summary_perfectly_slow_evolution_2})
play the same role as the perturbation $\delta H$ in Eq.~(\ref{eq:summary_evolution_operator_deltaH_perturbation_theory_first_order})
in the advanced theory. They slightly reshuffle the populations of
all instantaneous eigenstates in the system. This way, the exponential
suppression of the previous parts of the trajectory does not undermine
the ability of the end-point fastest growing eigenstate to become
dominant.

Therefore the explanation of Ref.~\citep{Zhang2019d} is correct.
Yet it is specific to non-smooth trajectories chosen deliberately,
whereas our advanced theory is applicable to general situations.

\subsection{Chiral dynamics without EP encircling}

\uline{Reference \mbox{\citep{Hassan2017b}} and its erratum \mbox{\citep{Hassan2017c}}}
have reported numerical and analytical results demonstrating chiral
conversion without encircling an EP. Further, experimental confirmation
of such behavior has been recently obtained \citep{Nasari2022}. We
have no reasons to doubt the validity of the numerical findings of
Refs.~\citep{Hassan2017b,Hassan2017c} and the experimental findings
of Ref.~\citep{Nasari2022}. However, we will now show that the analytical
considerations of Refs.~\citep{Hassan2017b,Hassan2017c} contain
a flaw and actually do not predict the phenomenon.

Reference \citep{Hassan2017b} employs the same Hamiltonian as in
our Eq.~(\ref{eq:nHH_examples}) an the trajectories that can be
described by our Eq.~(\ref{eq:trajectory_general}) with $\delta_{0}=0$,
$\phi=\pi/2$, and various values of $g_{0}$. That is, the same as
the trajectory considered in Sec.~\ref{subsec:chiral-without-encircling},
modulo the freedom of choosing the trajectory center, $g_{0}\in[0,1)$,
and radius, $R<1-g_{0}$. The authors use the Schrödinger equation,
$i\partial_{t}\ket{\psi}=H(t)\ket{\psi}$, for $\ket{\psi}=\left(a(t),b(t)\right)^{\mathrm{T}}$
to obtain
\begin{equation}
b(t)=a(t)\left(\delta(t)+ig(t)\right)-ia'(t),
\end{equation}
\begin{equation}
a''(t)+a(t)\left[1-g_{0}^{2}+e^{i\omega t}R\left(2g_{0}+i\gamma\right)-e^{2i\omega t}R^{2}\right]=0.\label{eq:Hassan_equation_for_a}
\end{equation}
Assuming $R$ small and neglecting the $R^{2}$ term, Eq.~(\ref{eq:Hassan_equation_for_a})
is then solved via
\begin{equation}
a(t)=C_{1}I_{\nu}(x_{0}e^{i\omega t/2})+C_{2}K_{\nu}(x_{0}e^{i\omega t/2}),\label{eq:Hassan_solution_for_a}
\end{equation}
where $I_{\nu}$ and $K_{\nu}$ are the modified Bessel functions,
$\nu=2\omega^{-1}\sqrt{1-g_{0}^{2}}$, $x_{0}=s\omega^{-1}\sqrt{R(2g_{0}+i\omega)}$,
and $C_{1}$, $C_{2}$ are arbitrary constants. Then considering $t=2\pi/\omega$
and taking the limit of $\omega\rightarrow0$, Ref.~\ref{eq:Hassan_equation_for_a}
derives the prediction of conversion to one of the eigenstates for
$\omega>0$, and to the other one for $\omega<0$, which constitutes
a prediction for chiral state conversion.

Note that this consideration concerns perfectly-controllable slow
evolution, and no fast noise. Therefore, our naïve theory would predict
no conversion whatsoever as $\mathrm{Im}\int_{0}^{T}dt\,\lambda_{\pm}(t)=0$,
cf.~Sec.~\ref{subsec:chiral-without-encircling}. Our advanced theory
would predict chiral conversion --- but only due to the uncontrollable
fast noise. We resolve this contradiction between our theory and the
analysis of Refs.~\citep{Hassan2017b,Hassan2017c} by showing that
Eq.~(\ref{eq:Hassan_solution_for_a}) does not actually predict conversion
in the limit $\omega\rightarrow0$.

To this end, we rewrite Eq.~(\ref{eq:Hassan_solution_for_a}) as
\begin{equation}
a(t)=AJ_{\nu}(x_{0}e^{i\omega t/2})+BJ_{-\nu}(x_{0}e^{i\omega t/2}),\label{eq:Hassan_solution_for_a_rewritten}
\end{equation}
where $J_{\nu}$ is the Bessel function of the first kind, and $A$,
$B$ are arbitrary constants. Note that the argumets of the Bessel
functions return to their original values after two evolution periods,
$t\rightarrow t+4\pi/\omega$. For a function without branch cuts,
this would immediately imply the absence of any amplification, as
the dynamics is periodic. However, $J_{\nu}(z)$ has a branch cut
$z\in(-\infty,0)$. This means that the amplification, if any, comes
from the mismatch on the branch cut. Using the Taylor expansion series,
one can express the Bessel function as
\begin{equation}
J_{\nu}(z)=\left(z/2\right)^{\nu}\,_{0}F_{1}(\nu+1;-z^{2}/4)/\Gamma(\nu+1),
\end{equation}
where $\,_{0}F_{1}$ is a generalized hypergeometric function. This
hypergeometric function, $\,_{0}F_{1}$, is an entire function, which
implies that it does not have any singularities for finite $z$. In
particular, no branch cuts. Therefore, the branch cut mismatch of
the Bessel function for $z<0$ comes from $z^{\nu}$ and is characterized
by $J_{\nu}(z+i0)/J_{\nu}(z-i0)=e^{2\pi i\nu}$. Given that $\nu\in\mathbb{R}$,
neither $J_{\nu}(x_{0}e^{i\omega t/2})$, nor $J_{-\nu}(x_{0}e^{i\omega t/2})$
undergoes amplification --- not only for $\omega\rightarrow0$, but
for any $\omega$. Further, even for the evolution of one time period,
$t\rightarrow t+2\pi/\omega$, the Bessel function changes only by
a phase:
\begin{multline}
J_{\pm\nu}(x_{0}e^{i\omega t/2})\rightarrow J_{\pm\nu}(x_{0}e^{i\omega t/2}e^{i\pi})\\
=e^{\pm i\pi\nu}J_{\pm\nu}(x_{0}e^{i\omega t/2}).
\end{multline}

One, therefore, finds that Eq.~(\ref{eq:Hassan_solution_for_a_rewritten})
and, equivalently, Eq.~(\ref{eq:Hassan_solution_for_a}) does not
predict amplification of one or another eigenstate, only non-trivial
coherent dynamics. This is in agreement with the prediction of our
theory for perfectly-controlled slow evolution. The erroneous analytical
result of Refs.~\citep{Hassan2017b,Hassan2017c} apparently stems
from a mistake when using the asymptotic expansions of $I_{\nu}$
and $K_{\nu}$ in the limit $\omega\rightarrow0$.

As for the numerical results of Refs.~\citep{Hassan2017b,Hassan2017c},
they can be understood in terms of our advanced theory incorporating
uncontrolled fast noise stemming from small numerical inaccuracies,
as discussed in Sec.~\ref{subsec:chiral-without-encircling}. We
believe that the same explanation by fast noise applies to the experimental
results of Ref.~\citep{Nasari2022}.

\section{Further demonstration of quantitative accuracy of the advanced theory}

\label{app:quantitative_advanced_theory}

In this Appendix we demonstrate that the advanced theory of Sec.~\ref{subsec:Summary_Slow-evolution-with-fast-noise}
can provide \emph{quantitatively} accurate predictions for non-Hermitian
evolution on the closed parameter space trajectories. We discuss examples
without (Sec.~\ref{subsec:Closed-evolution-no-EP}) and with (Sec.~\ref{subsec:Closed-evolution-with-EP})
encircling an EP. In all the examples, the evolution is generated
by the perturbed Hamiltonian (\ref{eq:nHH_examples_perturbed}), whose
unperturbed part is described with Eqs.~(\ref{eq:nHH_examples}--\ref{eq:trajectory_general})
and the perturbation is described by Eq.~(\ref{eq:deltaH_expression})
with $\varepsilon=10^{-4}$ and $\Omega=2\pi/5$. The need for explicitly
including a controlled perturbation was explained in Sec.~\ref{sec:quantitative_predictions_advanced_theory}.

\begin{widetext}

\subsection{\label{subsec:Advanced_theory_numerics}How we produce the state
evolution curves with the advanced theory}

The advanced theory that takes fast perturbations into account is
defined by Eq.~(\ref{eq:summary_evolution_operator_deltaH_perturbation_theory_first_order})
together with Eqs.~(\ref{eq:summary_perfectly_slow_evolution_1}--\ref{eq:summary_perfectly_slow_evolution_2})
and should be supplemented by the expressions for the $W$ and $\tilde{\ensuremath{W}}$
operators as derived in Appendix~\ref{app:Derivation_of_analytical_results}.
For the sake of the reader, here we present explicit formulas we use
to produce the advanced theory predictions in Figs.~\ref{fig:open-trajectory-perturbed}--\ref{fig:EP-non-chiral}.

According to Eq.~(\ref{eq:summary_evolution_operator_deltaH_perturbation_theory_first_order}),
the system state
\begin{equation}
\ket{\psi(t)}=\mathcal{\bar{U}}(t,0)\ket{\psi(0)}=\mathcal{U}(T,0)\ket{\psi(0)}-i\varepsilon\int_{0}^{T}dt_{1}\mathcal{U}(T,t_{1})\delta H(t_{1})\mathcal{U}(t_{1},0)\ket{\psi(0)}.
\end{equation}
In turn, Eqs.~(\ref{eq:summary_perfectly_slow_evolution_1}--\ref{eq:summary_perfectly_slow_evolution_2})
give
\begin{multline}
U^{-1}(t_{2})\mathcal{U}(t_{2},t_{1})U(t_{1})=\left(\mathbb{I}+W(t_{2})\right)\exp\left(-i\int_{t_{1}}^{t_{2}}d\tau\,D(\tau)\right)\left(\mathbb{I}+\tilde{W}(t_{1})\right)\\
\approx\exp\left(-i\int_{t_{1}}^{t_{2}}d\tau\,D(\tau)\right)+W^{(1)}(t_{2})\exp\left(-i\int_{t_{1}}^{t_{2}}d\tau\,D(\tau)\right)-\exp\left(-i\int_{t_{1}}^{t_{2}}d\tau\,D(\tau)\right)W^{(1)}(t_{1}),
\end{multline}
where neglected the terms of $O\left(T^{-2}\right)$, after using
Eqs.~(\ref{eq:W_k}--\ref{eq:Wtilde_k}, \ref{eq:Vk}) and (\ref{eq:V1}--\ref{eq:W1_explicit}).

Switching to the instantaneous eigenbasis $\ket{\varphi(t)}=U^{-1}(t)\ket{\psi(t)}$,
combining the above, and neglecting the terms of $O\left(T^{-2}\right)$,
one gets
\begin{multline}
\ket{\varphi(t)}=\underbrace{\left(\exp\left(-i\int_{0}^{t}d\tau\,D(\tau)\right)+W^{(1)}(t)\exp\left(-i\int_{0}^{t}d\tau\,D(\tau)\right)-\exp\left(-i\int_{0}^{t}d\tau\,D(\tau)\right)W^{(1)}(0)\right)\ket{\varphi(0)}}_{\text{term 1}}\\
\underbrace{-i\varepsilon\int_{0}^{t}dt_{1}\exp\left(-i\int_{t_{1}}^{t}d\tau\,D(\tau)\right)\delta\tilde{H}(t_{1})\exp\left(-i\int_{0}^{t_{1}}d\tau\,D(\tau)\right)\ket{\varphi(0)}}_{\text{term 2}}\\
\underbrace{-i\varepsilon W^{(1)}(t)\int_{0}^{t}dt_{1}\exp\left(-i\int_{t_{1}}^{t}d\tau\,D(\tau)\right)\delta\tilde{H}(t_{1})\exp\left(-i\int_{0}^{t_{1}}d\tau\,D(\tau)\right)\ket{\varphi(0)}}_{\text{term 3}}\\
\underbrace{+i\varepsilon\int_{0}^{t}dt_{1}\exp\left(-i\int_{t_{1}}^{t}d\tau\,D(\tau)\right)\delta\tilde{H}(t_{1})\exp\left(-i\int_{0}^{t_{1}}d\tau\,D(\tau)\right)W^{(1)}(0)\ket{\varphi(0)}}_{\text{term 4}}\\
\underbrace{+i\varepsilon\int_{0}^{t}dt_{1}\exp\left(-i\int_{t_{1}}^{t}d\tau\,D(\tau)\right)\left[W^{(1)}(t_{1}),\delta\tilde{H}(t_{1})\right]\exp\left(-i\int_{0}^{t_{1}}d\tau\,D(\tau)\right)\ket{\varphi(0)}}_{\text{term 5}},\label{eq:advanced_simulation_formula}
\end{multline}
where $\delta\tilde{H}(t_{1})=U^{-1}(t_{1})\delta H(t_{1})U(t_{1})$
is the perturbation in the instantaneous eigenbasis and
\begin{equation}
\left[W^{(1)}(t_{1}),\delta\tilde{H}(t_{1})\right]=W^{(1)}(t_{1})\delta\tilde{H}(t_{1})-\delta\tilde{H}(t_{1})W^{(1)}(t_{1}).
\end{equation}
The expression for $W^{(1)}(t)$ should be evaluated using Eq.~(\ref{eq:W1_explicit}).
The term numbering in Eq.~(\ref{eq:advanced_simulation_formula})
corresponds to their labeling in our code \citep{GitHubRepository}
that was used to produce Figs.~\ref{fig:open-trajectory-perturbed}--\ref{fig:EP-non-chiral}.

\end{widetext}

\subsection{\label{subsec:Closed-evolution-no-EP}Closed evolution: without encircling
the exceptional point}

\begin{figure*}[p]
\begin{centering}
\includegraphics[width=1\textwidth]{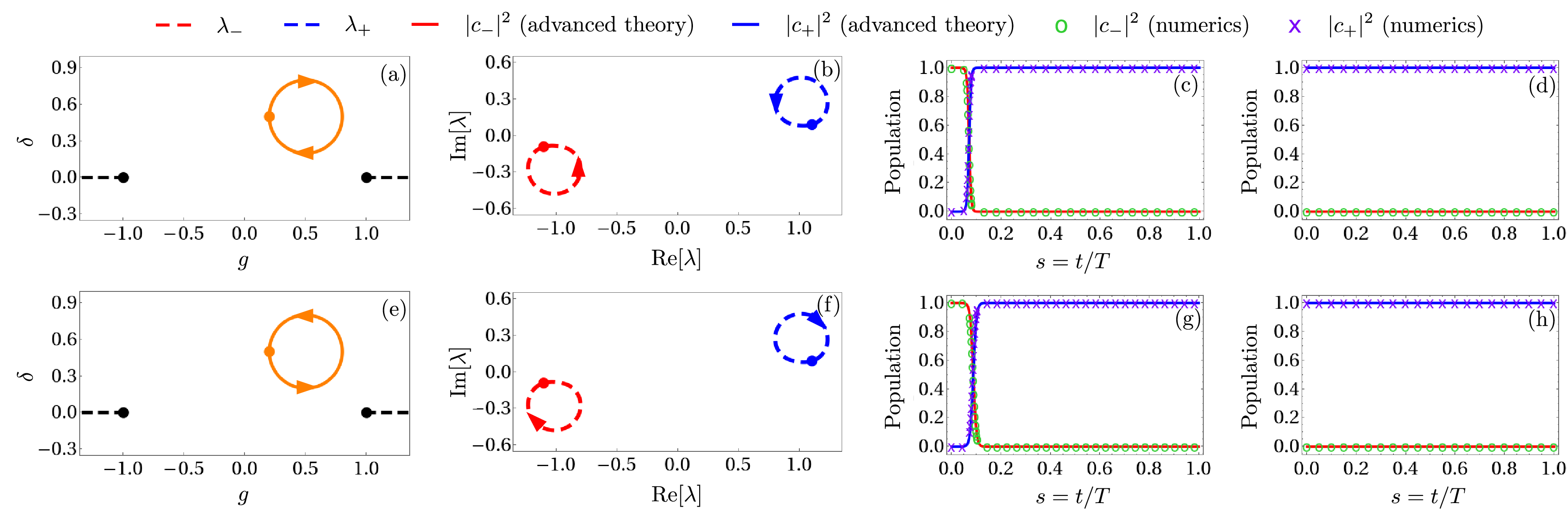}
\par\end{centering}
\caption{\textbf{Closed trajectory }\textbf{\emph{not encircling an EP}}\textbf{
can lead to }\textbf{\emph{non-chiral}}\textbf{ state conversion.}
(a) and (e) --- The clockwise and counterclockwise trajectories (orange)
correspond to the following parameters in Eq.~(\ref{eq:trajectory_general}):
$\delta_{0}=0.5$, $g_{0}=0.5$, $R=0.3$, $T=500$, $\omega=-2\pi/T$,
$\phi=0$ (clockwise); $\delta_{0}=0.5$, $g_{0}=0.5$, $R=0.3$,
$T=500$, $\omega=2\pi/T$, $\phi=0$ (counterclockwise). The trajectory
direction is shown with arrows and the orange dot depict the starting
point. The black dots correspond to the EPs of the Hamiltonian, and
the dashed lines correspond to the branch cuts of $\lambda_{\pm}$.
(b) and (f) --- The trajectory of the eigenvalues $\lambda_{\pm}$
corresponding to the respective trajectory. The effect of perturbation
$\varepsilon\delta H(t)$, cf.~Eqs.~(\ref{eq:nHH_examples_perturbed}--\ref{eq:deltaH_expression})
is clearly not noticeable here. The eigevalues $\lambda_{\pm}$ at
the start of the parameter trajectory are depicted as blue and red
dots, and the arrows show the trajectory direction. (c) and (g) ---
The population dynamics for the respective trajectory when the system
is initialized in $\protect\ket{\psi_{-}}$. (d) and (h) --- The
population dynamics for the respective trajectory encircling when
the system is initialized in $\protect\ket{\psi_{+}}$.}

\label{fig:closed-no-EP-nonchiral}
\end{figure*}

The evolution not encircling any exceptional points can generate both
non-chiral and chiral state conversion behaviors. We start with a
\emph{non-chiral example}. As can be seen in Fig.~\ref{fig:closed-no-EP-nonchiral},
irrespective of the trajectory direction, essentially any initial
state is converted to $\ket{\psi_{+}}$. This is easy to understand
since always $\mathrm{Im}\,\lambda_{+}>0>\mathrm{Im}\,\lambda_{-}$,
and therefore $\ket{\psi_{+}}$ is always the end-point fastest growing
state. The theoretical population dynamics calculated according to
the advanced theory is in quantitative agreement with the numerical
simulations.

\begin{figure*}[p]
\begin{centering}
\includegraphics[width=1\textwidth]{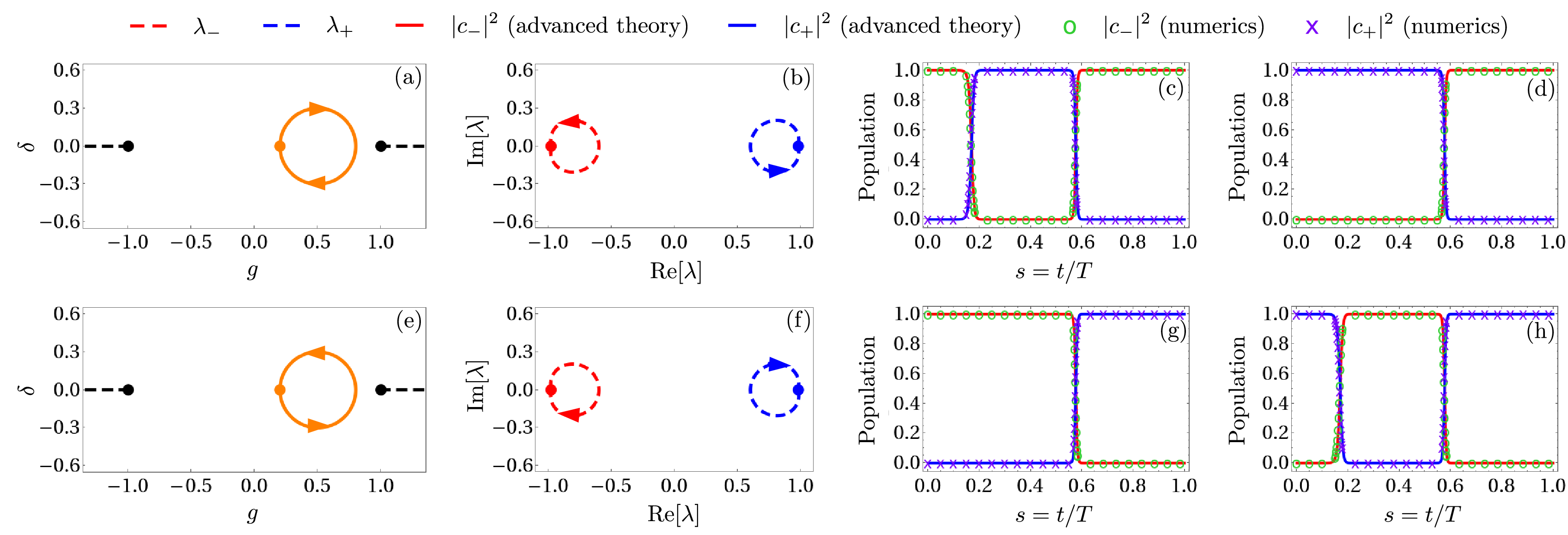}
\par\end{centering}
\caption{\textbf{Closed trajectory }\textbf{\emph{not encircling an EP}}\textbf{
can lead to }\textbf{\emph{chiral}}\textbf{ state conversion.} (a)
and (e) --- The clockwise and counterclockwise trajectories (orange)
correspond to the following parameters in Eq.~(\ref{eq:trajectory_general}):
$\delta_{0}=0$, $g_{0}=0.5$, $R=0.3$, $T=500$, $\omega=-2\pi/T$,
$\phi=0$ (clockwise); $\delta_{0}=0$, $g_{0}=0.5$, $R=0.3$, $T=500$,
$\omega=2\pi/T$, $\phi=0$ (counterclockwise). The trajectory direction
is shown with arrows and the orange dot depict the starting point.
The black dots correspond to the EPs of the Hamiltonian, and the dashed
lines correspond to the branch cuts of $\lambda_{\pm}$. (b) and (f)
--- The trajectory of the eigenvalues $\lambda_{\pm}$ corresponding
to the respective trajectory. The effect of perturbation $\varepsilon\delta H(t)$,
cf.~Eqs.~(\ref{eq:nHH_examples_perturbed}--\ref{eq:deltaH_expression})
is clearly not noticeable here. The eigevalues $\lambda_{\pm}$ at
the start of the parameter trajectory are depicted as blue and red
dots, and the arrows show the trajectory direction. (c) and (g) ---
The population dynamics for the respective trajectory when the system
is initialized in $\protect\ket{\psi_{-}}$. (d) and (h) --- The
population dynamics for the respective trajectory encircling when
the system is initialized in $\protect\ket{\psi_{+}}$.}

\label{fig:closed-no-EP-chiral}
\end{figure*}

For a different trajectory, cf.~Fig.~\ref{fig:closed-no-EP-chiral},
the conversion is \emph{chiral}. A clockwise trajectory results in
conversion to $\ket{\psi_{-}}$, while following the same trajectory
in the counterclockwise direction results in conversion to $\ket{\psi_{+}}$.
In fact, this is the same trajectory that was considered in Sec.~\ref{subsec:chiral-without-encircling}.
As it was explained there, the conversion is to the end-point fastest
growing state. Comparing the numerical results in Fig.~\ref{fig:closed-no-EP-chiral}
and Fig.~\ref{fig:chiral-without-encircling}, one can clearly see
the effect of perturbation $\varepsilon\delta H(t)$ (\ref{eq:deltaH_expression})
that is present in Fig.~\ref{fig:closed-no-EP-chiral} and absent
in Fig.~\ref{fig:chiral-without-encircling}. This shows that $\varepsilon\delta H(t)$
is the dominant pertrubation in the system, overshadowing the numerical
errors. The theoretical population dynamics calculated according to
the advanced theory is in quantitative agreement with the numerical
simulations, as can be seen in Figs.~\ref{fig:closed-no-EP-chiral}(c,d,g,h).

\subsection{\label{subsec:Closed-evolution-with-EP}Closed evolution: with encircling
the exceptional point}

\begin{figure*}[p]
\begin{centering}
\includegraphics[width=1\textwidth]{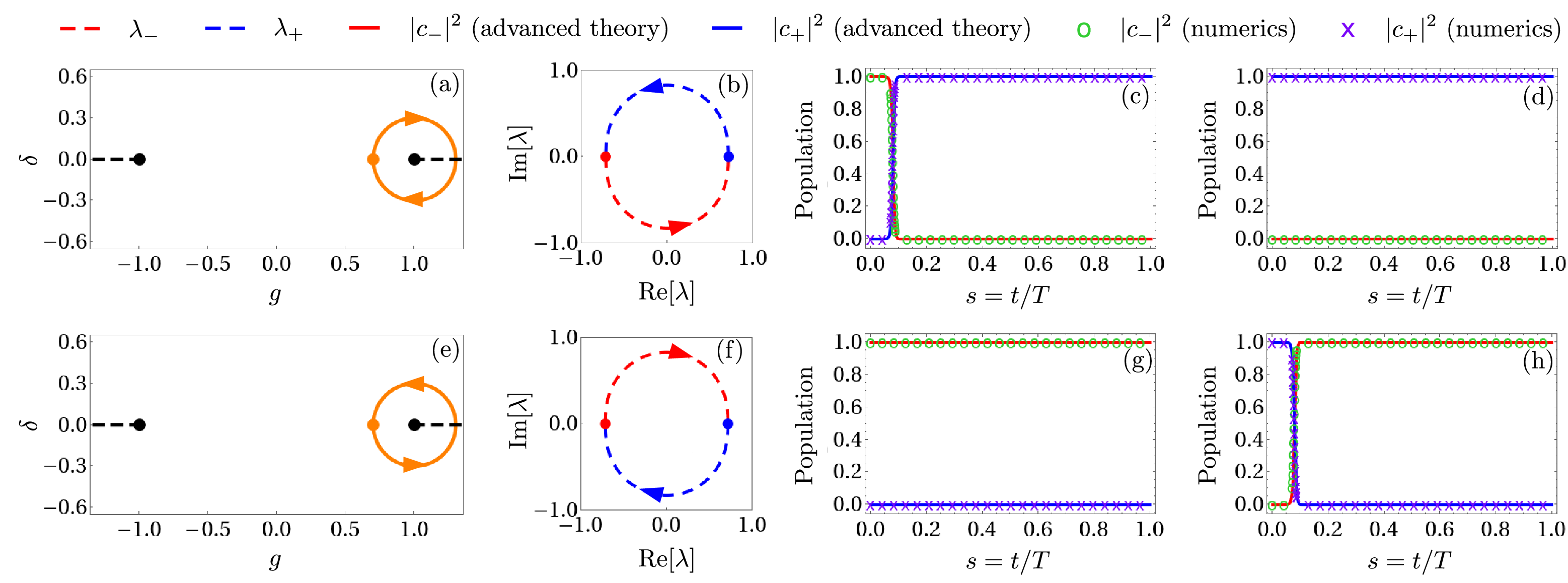}
\par\end{centering}
\caption{\textbf{Closed trajectory}\textbf{\emph{ encircling an EP}}\textbf{
can lead to }\textbf{\emph{chiral}}\textbf{ state conversion.} (a)
and (e) --- The clockwise and counterclockwise trajectories (orange)
correspond to the following parameters in Eq.~(\ref{eq:trajectory_general}):
$\delta_{0}=0$, $g_{0}=1$, $R=0.3$, $T=500$, $\omega=-2\pi/T$,
$\phi=0$ (clockwise), $\delta_{0}=0$, $g_{0}=1$, $R=0.3$, $T=500$,
$\omega=2\pi/T$, $\phi=0$ (counterclockwise). The trajectory direction
is shown with arrows and the orange dot depict the starting point.
The black dots correspond to the EPs of the Hamiltonian, and the dashed
lines correspond to the branch cuts of $\lambda_{\pm}$. (b) and (f)
--- The trajectory of the eigenvalues $\lambda_{\pm}$ corresponding
to the respective trajectory. The effect of perturbation $\varepsilon\delta H(t)$,
cf.~Eqs.~(\ref{eq:nHH_examples_perturbed}--\ref{eq:deltaH_expression})
is clearly not noticeable here. The eigevalues $\lambda_{\pm}$ at
the start of the parameter trajectory are depicted as blue and red
dots, and the arrows show the trajectory direction. (c) and (g) ---
The population dynamics for the respective trajectory when the system
is initialized in $\protect\ket{\psi_{-}}$. (d) and (h) --- The
population dynamics for the respective trajectory encircling when
the system is initialized in $\protect\ket{\psi_{+}}$.}
\label{fig:EP-chiral}
\end{figure*}

The evolution encircling an EP can similarly generate both non-chiral
and chiral state conversion behaviors. The more conventional \emph{chiral}
conversion upon encircling an EP is shown in Fig.~\ref{fig:EP-chiral}.
Panels (c--d) and (g--h) show quantitative agreement between the
numerical simulations and the predictions of the advanced theory.

\begin{figure*}[p]
\begin{centering}
\includegraphics[width=1\textwidth]{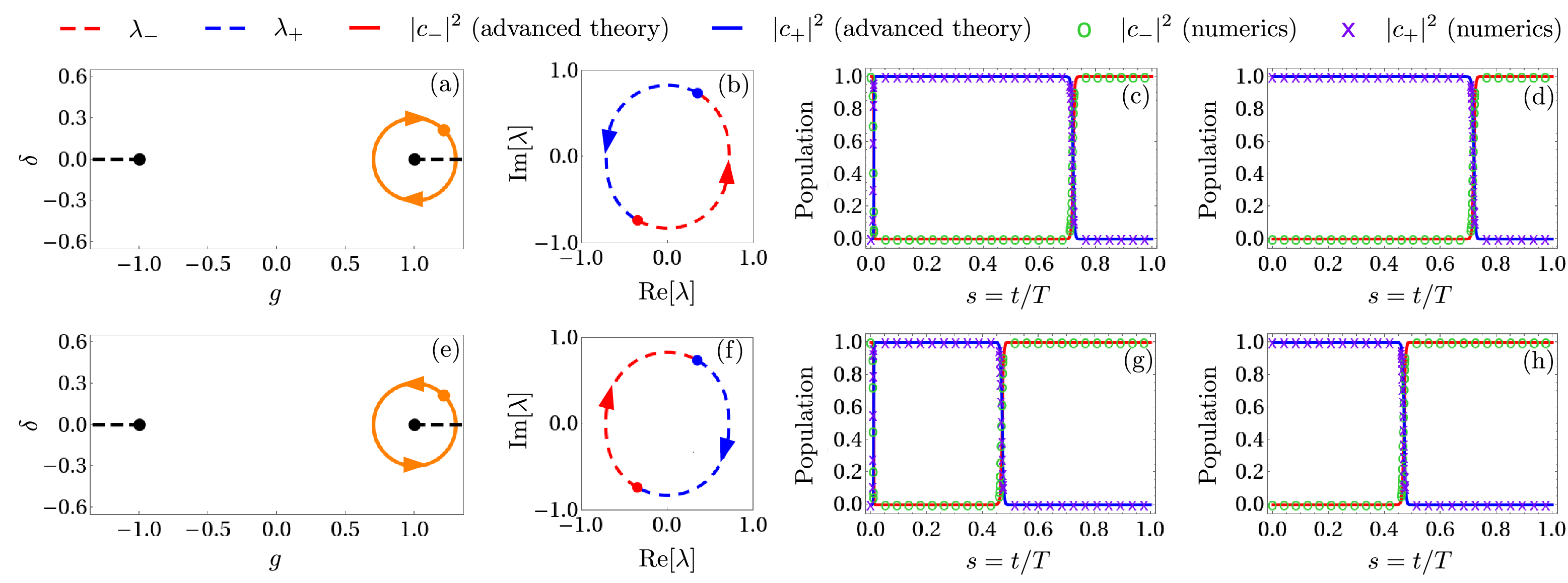}
\par\end{centering}
\caption{\textbf{Closed trajectory}\textbf{\emph{ encircling an EP}}\textbf{
can lead to }\textbf{\emph{non-chiral}}\textbf{ state conversion.}
(a) and (e) --- The clockwise and counterclockwise trajectories (orange)
correspond to the following parameters in Eq.~(\ref{eq:trajectory_general}):
$\delta_{0}=0$, $g_{0}=1$, $R=0.3$, $T=500$, $\omega=-2\pi/T$,
$\phi=-3\pi/4$ (clockwise), $\delta_{0}=0$, $g_{0}=1$, $R=0.3$,
$T=500$, $\omega=2\pi/T$, $\phi=-3\pi/4$ (counterclockwise). The
trajectory direction is shown with arrows and the orange dot depict
the starting point. The black dots correspond to the EPs of the Hamiltonian,
and the dashed lines correspond to the branch cuts of $\lambda_{\pm}$.
(b) and (f) --- The trajectory of the eigenvalues $\lambda_{\pm}$
corresponding to the respective trajectory. The effect of perturbation
$\varepsilon\delta H(t)$, cf.~Eqs.~(\ref{eq:nHH_examples_perturbed}--\ref{eq:deltaH_expression})
is clearly not noticeable here. The eigevalues $\lambda_{\pm}$ at
the start of the parameter trajectory are depicted as blue and red
dots, and the arrows show the trajectory direction. (c) and (g) ---
The population dynamics for the respective trajectory when the system
is initialized in $\protect\ket{\psi_{-}}$. (d) and (h) --- The
population dynamics for the respective trajectory encircling when
the system is initialized in $\protect\ket{\psi_{+}}$.}

\label{fig:EP-non-chiral}
\end{figure*}

The \emph{non-chiral} conversion is exemplified in Fig.~\ref{fig:EP-non-chiral}.
The parameter trajectory is the same as in Sec.~\ref{subsec:non-chiral-with-encircling}.
Comparing the times of population switches between Figs.~\ref{fig:EP-non-chiral}
and \ref{fig:non-chiral-with-encircling}, one sees that, again, here
the controlled perturbation $\varepsilon\delta H(t)$ (\ref{eq:deltaH_expression})
overshadows the perturbations due to numerical errors. Under these
conditions, panels (c--d) and (g--h) show quantitative agreement
between the numerical simulations and the predictions of the advanced
theory. Note that the times of last population switch are not the
same for the clockwise and the counterclockwise trajectories. This
is because $\mathrm{Im}\,\lambda_{-}$ becomes larger than $\mathrm{Im}\,\lambda_{+}$
at different times depending on the trajectory direction. This further
illustrates that it is the end-point fastest growing state that wins
--- the overall growth/decay over the early parts of the trajectory
does not matter.

\end{document}